\documentclass{bmcart}

\usepackage{amsthm,amsmath}
\usepackage{hyperref}
\usepackage[utf8]{inputenc} 
\usepackage{graphicx}

\startlocaldefs

\DeclareGraphicsExtensions{.pdf,.jpg,.png}
\graphicspath{ {./Fig/} }

\renewcommand{\theenumi}{\roman{enumi}}

\newenvironment{enum}
{\begin{enumerate}
  \setlength{\itemsep}{2pt}
  \setlength{\parskip}{0pt}
  \setlength{\parsep}{0pt}}
{\end{enumerate}}

\def\et{\epsilon_T}
\def\mt{\mu_T}
\def\ea{\epsilon_A}
\def\eac{\epsilon_A^*}
\def\ma{\mu_A}
\def\pn{p_{\text{new}}}
\def\tmax{\tau_{\text{max}}}
\def\I{\mathbf{I}}
\def\II{\mathbf{II}}
\def\III{\mathbf{III}}

\endlocaldefs

\begin{document}

\begin{frontmatter}

\begin{fmbox}
\dochead{Research}


\title{Modeling Social Dynamics in a Collaborative Environment}

  
\author[
   addressref={aff_aalto}, 
   noteref={n_cont}, 
   email={gerardo.iniguez@aalto.fi} 
]{\inits{G}\fnm{Gerardo} \snm{I\~niguez}}
\author[
   addressref={aff_bute},
   noteref={n_cont}
]{\inits{J}\fnm{J\'anos} \snm{T\"or\"ok}}
\author[
   addressref={aff_oii},
   corref={aff_oii}, 
   email={taha.yasseri@oii.ox.ac.uk}
]{\inits{T}\fnm{Taha} \snm{Yasseri}}
\author[
   addressref={aff_aalto}
]{\inits{K}\fnm{Kimmo} \snm{Kaski}}
\author[
   addressref={aff_ceu,aff_bute}
]{\inits{J}\fnm{J\'anos} \snm{Kert\'esz}}


\address[id=aff_aalto]{ 
  \orgname{Department of Biomedical Engineering and Computational Science, Aalto University School of Science}, 
  \postcode{FI-00076}, 
  \city{Aalto}, 
  \cny{Finland} 
}

\address[id=aff_bute]{
  \orgname{Institute of Physics, Budapest University of Technology and Economics},
  \postcode{H-1111},
  \city{Budapest},
  \cny{Hungary}
}

\address[id=aff_oii]{
  \orgname{Oxford Internet Institute, University of Oxford},
  \postcode{OX1 3JS},
  \city{Oxford},
  \cny{United Kingdom}
}

\address[id=aff_ceu]{
  \orgname{Center for Network Science, Central European University},
  \postcode{H-1051},
  \city{Budapest},
  \cny{Hungary}
}


\begin{artnotes}
\note[id=n_cont]{Equal contributors} 
\end{artnotes}

\end{fmbox}


\begin{abstractbox}

\begin{abstract} 
Wikipedia is a prime example of today’s value production in a
collaborative environment. Using this example, we model the emergence,
persistence and resolution of severe conflicts during collaboration by
coupling opinion formation with article editing in a bounded
confidence dynamics. The complex social behavior involved in editing articles is implemented as a minimal model with two basic elements;
(i) individuals interact directly to share information and convince each other, and
(ii) they edit a common medium to establish their own opinions.
Opinions of the editors and that represented by the article are
characterised by a scalar variable. When the pool of editors is fixed,
three regimes can be distinguished: (a) a stable mainstream article
opinion is continuously contested by editors with extremist views and there is
slow convergence towards consensus, (b) the article oscillates
between editors with extremist views, reaching consensus relatively fast
at one of the extremes, and (c) the extremist editors are
converted very fast to the mainstream opinion and the article has an erratic evolution.
When editors are renewed with a certain rate, a dynamical
transition occurs between different kinds of edit wars, which
qualitatively reflect the dynamics of conflicts as observed in real
Wikipedia data.

\end{abstract}



\end{abstractbox}
%

\end{frontmatter}




\section{Introduction}
\label{sec:intro}

Cooperative societies are ubiquitous in nature~\cite{axelrod1981evolution}, yet the cooperation or the mutual assistance between members 
of a society is also likely to generate conflicts~\cite{schelling1980strategy}.  Potential for conflicts is commonplace even in insect 
species~\cite{ratnieks2006conflict} and so is conflict management through policing and negotiation in groups of 
primates~\cite{waal2000primates,flack2006policing}. In human societies cooperation goes further not only in its scale and 
range, but also in the available mechanisms to promote conflict resolution~\cite{melis2010human,rand2011dynamic}. 
Collaborative and conflict-prone human endeavors are numerous, including public policy-making in globalized 
societies~\cite{quirk1989cooperative,buchan2009globalization}, open-source software development~\cite{lerner2002some}, 
teamwork in operating rooms~\cite{rogers2011teaching}, and even long-term partnerships~\cite{minson2011two}. Moreover, 
information communication technology opens up entirely new ways of collaboration.
With such a diversity in system size and social interactions between individuals, it seems appropriate to study this phenomenon 
of social dynamics in the framework of statistical physics~\cite{castellano2009statistical,helbing2010quantitative}, an approach 
benefiting greatly from the availability of large scale data on social interactions~\cite{onnela2007structure,ratkiewicz2010characterizing}.

As a relevant example of conflicts in social cooperation we select Wikipedia (WP), an intriguing example of value 
production in an online collaborative environment~\cite{yasseri2013value}. WP is a free web-based encyclopedia
project, where volunteering individuals collaboratively write and edit articles about any topic they choose. 
The availability of virtually all data concerning the visiting and editing of article pages provides a solid 
empirical basis for investigating topics such as online content popularity~\cite{ratkiewicz2010characterizing,mestyan2013} and
the role of opinion-formation processes in a peer-production environment~\cite{ciampaglia2011bounded}. 

The editing process in WP is usually peaceful and constructive, but some controversial topics might give rise to extreme cases 
of disagreement over the contents of the articles, with the editors repeatedly overriding each other's contributions and 
making it harder to reach consensus. These `edit wars' (as they are commonly called) result from complex online social dynamics, and recent 
studies~\cite{yasseri2012dynamics} have shown how to detect and classify them, as well as how they are related to 
burstiness and what are their circadian patterns in editing activity~\cite{yasseri2012circadian}. 

Although collaborative behavior might appear without direct interactions between individuals, communication tends to have a positive effect on cooperation and trust~\cite{kollock1998social}. Indeed, more immediate forms of communication (voice as opposed to text, for example) have been seen to increase the level of cooperation in online environments~\cite{jensen2000effect}. In WP, direct communication is implemented via `talk pages', open forums where editors may discuss improvements over the contents of articles and exchange their related opinions~\cite{wiki:talk}. Discussions among editors are not mandatory~\cite{wiki:using_talk}, 
but there is a significant correlation between talk page length and the likelihood of an edit war, indicating that many debates happen in articles and talk pages, simultaneously~\cite{yasseri2013value,kaltenbrunner2012there}.

Overall, a minimal model aimed at reproducing the temporal evolution of a common medium (i.e. a product collectively modified by a group of people, like an article in WP) requires at least the following two ingredients:
\begin{enum}
\item \textit{agent-agent dynamics}: Individuals share their views and opinions about changes in the article 
using an open channel accessible to all editors (the talk page or some other means of communication), thus effectively participating in an 
opinion-formation process through information sharing.
\item \textit{agent-medium dynamics}: Individuals edit the article if it does not properly summarize their views 
on the subject, thus controlling the temporal evolution of the article and coupling it to the opinion-formation mechanism.
\end{enum}

We describe the opinion-formation process taking place in the talk page by means of the 
well-known \textit{bounded confidence} mechanism first introduced by Deffuant \textit{et al.}~\cite{deffuant2000mixing}, 
where real discussions take place only if the opinions of the people involved are sufficiently close to each other. 
Conversely, we model article editing by an `inverse' bounded confidence process, where individuals change the current 
state of the article only if it differs too much from their own opinions. Particularly, we focus our attention on how 
the coupling between agent-agent and agent-medium interactions determines the nature of the temporal evolution of an article. 
This we consider as a further step towards the theoretical characterization of conflict in social cooperative systems 
such as WP~\cite{torok2013opinions}.

The text is organized as follows: In Section~\ref{sec:model} we introduce and discuss the model in detail. In Section~\ref{sec:results} 
we describe our results separately for the cases of a fixed editor pool and a pool with editor renewal, and finally make a comparison with 
empirical observations on WP conflicts. In Section~\ref{sec:conc} we present concluding remarks.

\section{Model}
\label{sec:model}

Let us first assume that there is a system of $N$ agents as potential editors for a collective medium. The state 
of an individual $i$ at time $t$ is defined by its scalar, continuous opinion $x_i(t) \in [0, 1]$, while the medium 
is characterized by a certain value $A(t)$ in that same interval. The variable $x$ represents the view and/or 
inclination of an agent concerning the topic described by the common medium, while $A$ is the particular position 
actually represented by the medium.

Although it may seem too reductive to describe people's perceptions by a scalar variable $x$, many topics can actually be projected to a one-dimensional struggle between two extreme, opposite options. In the Liancourt Rocks territorial dispute between South Korea and Japan~\cite{liancourt:dispute}, for example, the values $x = 0,1$ represent the extreme position of favoring sovereignty of the islets for a particular country. Other topics are of course multifaceted, generating discussions that depend on the global affinity of individuals and multiple cultural factors~\cite{axelrod1997dissemination}. While this complexity could be tackled by the use of vectorial opinions~\cite{lorenz2007continuous,sznajd2005left}, our intention here is not to describe extremism as realistically as possible, but to study the rise of collaborative conflict even in the simplest case of binary extremism.

In the case of WP, the scalar variable $A$ represents the opinion expressed by the written contents of an article, which carries the assumption that all agents perceive the medium in the same way. Real scenarios of public opinion might be more complex, given the tendency of individuals to attribute their own views to others and thus perceive false consensus~\cite{wojcieszak2009underlies}, usually out of a social need to belong~\cite{morrison2011socially}. Even so, we consider $A$ to be a sensible description of a WP article, one that could initially be measured by human judgment in the form of expert opinions, or in an automated way by quantifying lexical features and the use of certain language constructs. We note, however, that the actual value of $A$ is not the main concern of our study. Instead, we are interested in how opinion differences in collaborative groups may eventually lead to conflict, specifically when such opinion differences are perceived with respect to a common medium that all 
individuals modify collectively.

\subsection{Agent-agent dynamics}
\label{ssec:AAD}

For the agent-agent dynamics (AAD) we consider a generic bounded-confidence model over a complete 
graph~\cite{deffuant2000mixing,weisbuch2002meet}, that is, a succession of random binary encounters among 
all agents in the system. We initialize every opinion $x_i(0)$ to a uniformly-distributed random value in 
the interval $[0, 1]$. The initial medium value $A(0)$ is chosen uniformly at random from the same interval. 
This way, even an initially moderate medium $A \sim 1/2$ may find discord with extreme opinions at the boundaries. 
For each interaction we randomly select two agents $i$, $j$ and compare their opinions. If the difference in 
opinions exceeds a given threshold $\et$ nothing happens, but if $|x_i - x_j| < \et$ we update as follows,
\begin{equation}
\label{eq:deffuant}
(x_i, x_j) \mapsto (x_i + \mt [ x_j - x_i ], x_j + \mt [ x_i - x_j ]).
\end{equation}
The parameter $\et \in [0, 1]$ is usually referred to as the \textit{confidence} or \textit{tolerance} for 
pairwise interactions, while $\mt \in [0, 1/2]$ is a \textit{convergence} parameter. AAD is then a symmetric 
compromise between similarly-minded individuals: people with very different opinions simply do not pay attention 
to each other, but similar agents debate and converge their views by the relative amount $\mt$.

The dynamics set by Eq.~(\ref{eq:deffuant}) has received a lot of attention in the past~\cite{castellano2009statistical}, 
starting from the mean-field description of two-body inelastic collisions in granular gases~\cite{ben2000multiscaling,baldassarri2002influence}. 
Its final, steady state is comprised by $n_c \sim 1/(2\et)$ isolated opinion groups that arise due to the instability of the initial 
opinion distribution near the boundaries. Furthermore, $n_c$ increases as $\et \to 0$ in a series of bifurcations~\cite{ben2003bifurcations}. 
In the limit $\mt \to 0$ corresponding to a `stubborn' society, the asymptotically final value of $n_c$ also depends 
on $\mt$~\cite{laguna2004minorities,porfiri2007decline}. The bounded-confidence mechanism has been extended in many ways over the years, 
considering interactions between more than two agents~\cite{hegselmann2002opinion}, vectorial opinions~\cite{lorenz2007continuous,fortunato2005vector,jacobmeier2005multidimensional,lorenz2008fostering}, 
and coupling with a constant external field~\cite{gonzalez2010spontaneous}.

\subsection{Agent-medium dynamics}
\label{ssec:AMD}

For the agent-medium dynamics (AMD) we use what could be thought of as an asymmetric, inverse version of the bounded-confidence mechanism 
described above. When the opinion of a randomly chosen agent $i$ is very different from the current 
state of the medium, namely if $|x_i - A| > \ea$, we make the update,
\begin{equation}
\label{eq:article}
A \mapsto A + \ma [ x_i - A ],
\end{equation}
where $\ea$, $\ma \in [0, 1]$ are the tolerance and convergence parameters for AMD. In other words, individuals that 
come across a version of the medium portraying a radically different set of mind will modify it by the relative 
amount $\ma$, where the threshold to define similarity is given by $\ea$. Conversely, if $|x_i - A| < \ea$ we update,
\begin{equation}
\label{eq:opinion}
x_i \mapsto x_i + \ma [ A - x_i ].
\end{equation}
meaning that individuals edit the medium when it differs too much from their opinions, but adopt the medium's 
view when they already think similarly. Observe that the maximum meaningful value of $\mt$ is 1/2 (i.e. convergence 
to the average of opinions), while the maximum $\ma = 1$ implies changing the medium (opinion) so that it completely 
reflects the agent's (medium's) point of view.

The previous rules comprise our model for the dynamics of conflicts in WP given a \textit{fixed agent pool}, 
that is, without agents entering or leaving the editing process of the common medium. In a numerical implementation 
of the model, every time step $t$ consists of $N$ updates of AAD given by Eq.~(\ref{eq:deffuant}) and of AMD 
following Eqs.~(\ref{eq:article}) and~(\ref{eq:opinion}), so that time is effectively measured in number of edits 
and the broad inter-event time distribution between successive edits (observed in empirical studies~\cite{yasseri2012dynamics})
does not have to be considered directly. Given a fixed agent pool, AAD favors opinion homogenization in intervals of 
length $2\et$ and can thus create several opinion groups for low tolerance, while AMD makes the medium value follow the 
majority group. Then, for a finite system there is a nonzero probability that any agent outside the majority group 
will be drawn by the medium to it, and the system will always reach consensus after a transient regime characterized 
by fluctuations in the medium value~\cite{torok2013opinions}.

However, in real WP articles the pool of editors tends to change frequently. Some editors leave (due to 
boredom, lack of interest or fading media coverage on the subject, or are banned from editing by editors at a higher hierarchical level) 
and newly arrived agents do not necessarily share 
the opinions of their predecessors. Such feature of \textit{agent renewal} during the process or writing an article may 
destroy consensus and lead to a steady state of alternating conflict and consensus phases, which we take into account by 
introducing thermal noise in the model. Along with any update of AAD/AMD, one editor might leave the 
pool with probability $\pn$ and be substituted by a new agent with opinion chosen uniformly at random from the 
interval $[0, 1]$. The quantity $1/(N\pn)$ then formally acts as the inverse temperature of the system, signaling 
a dynamical phase transition between different regimes of conflict~\cite{torok2013opinions}.

\section{Results}
\label{sec:results}

\subsection{Fixed agent pool}
\label{ssec:fixedPool}

In the presence of a fixed agent pool ($\pn=0$) with finite size $N$, the dynamics always reaches a peaceful state where 
all agents' opinions lie within the tolerance of the medium. To show this, let us calculate the probability that 
an unsatisfied editor $i$ changes the medium $A$ for $n$ consecutive times, such that afterwards $|x_i - A'| < \ea$ and 
the agent can finally stop its stream of edits. For fixed $x_i$ and following Eq.~(\ref{eq:article}), the final 
distance between editor and medium is $|x_i - A'| = (1 - \ma)^n |x_i - A|$, so the inequality $|x_i - A'| < \ea$ is 
satisfied if $n > \ln\ea / \ln(1-\ma)$. The probability of agent $i$ not participating in AAD for $n$ time 
steps is $(1 - 2/N)^n$, while the probability of choosing it for AMD is $1/N^n$. Then the total probability of this 
stream of edits is $(1 - 2/N)^n / N^n$, which for large $N$ and $\ma \sim 0$ might be very small, but always finite. 
After editor $i$ gets into the tolerance interval of the medium, it will not perform additional edits and will 
eventually adopt the majority opinion close to the medium value. Similar events with other unsatisfied agents will 
finally result in full consensus and put an end to the dynamics.

The existence of a finite relaxation time $\tau$ to consensus (for finite systems) contrasts drastically with the behavior of the bounded confidence mechanism alone, where consensus is never attained for $\et < 1/2$~\cite{castellano2009statistical}. In other words, the presence of agent-medium interactions 
promotes an agreement of opinions that would otherwise not exist in the agent-agent dynamics, even though it may happen after a very long time (i.e. $\tau \gg 0$). If we think of the medium as an additional agent with maximum tolerance (in the sense that it always interacts with the rest no matter what) and against which agents have a different tolerance $\ea$ (as opposed to $\et$), this result is reminiscent of previous observations for a bounded-confidence model with heterogeneous thresholds~\cite{weisbuch2002meet,lorenz2010heterogeneous}. There, even a small fraction of `open-minded' agents with relatively high tolerance may bridge the opinion difference between the rest of the agents and lead to consensus.

In order to analyze all possible typical behaviors of the fixed agent pool dynamics, we perform extensive numerical simulations in systems of 
size ranging from $N = 10$ to $10^4$, letting the dynamics evolve for a maximum time $\tmax = 10^4$. We then characterize the temporal evolution 
of medium and agent opinions as a function of $\et$, $\ea$ and $\ma$, while keeping $\pn = 0$ for all results in this Section. Finally, since the 
value of $\mt$ has no major effect other than regulating the convergence time of AAD~\cite{laguna2004minorities,porfiri2007decline}, from now on 
we fix it to the maximum value $\mt = 1/2$ in order to speed up the simulations as much as possible.

\begin{figure}[ht!]
\includegraphics[width=0.95\linewidth]{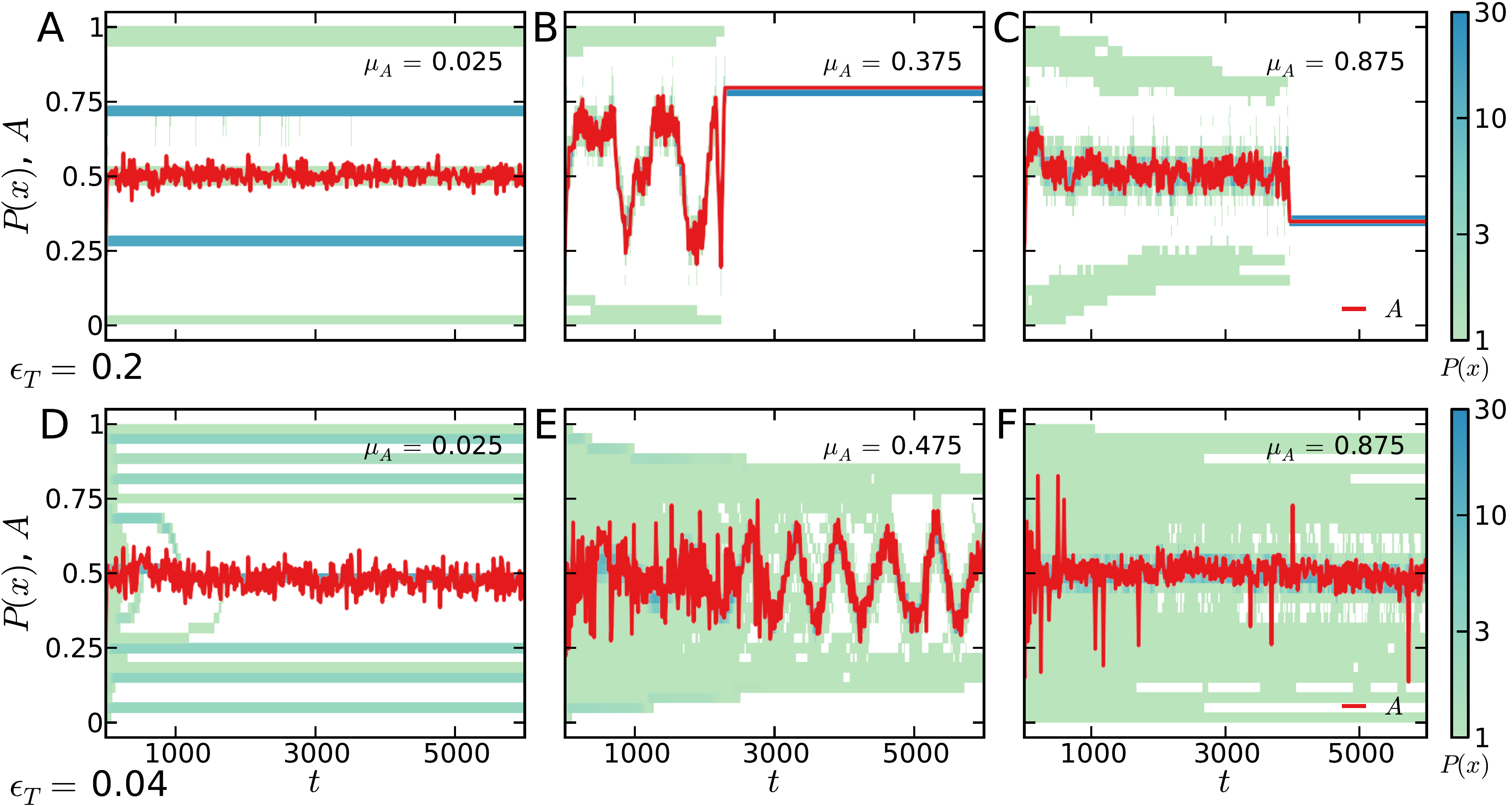}
\caption{\csentence{ Temporal evolution of opinions and medium.} (A, B, C) Time series of both the 
density distribution $P(x)$ of agents' opinions $x$ (color map) and the medium value $A$ (line) for $\et = 0.2$ and several $\ma$ 
values, signaling the three different regimes found in the dynamics. (D, E, F) The same but for $\et = 0.04$. 
Simulations correspond to $\ea = 0.1$ and $N = 10^4$.}
\label{fig:tseries_figure}
\end{figure}

A sample time series of medium and agent opinions is shown in Fig.~\ref{fig:tseries_figure}. As a function of the medium convergence $\ma$ the 
temporal evolution of the system shows three distinctive behaviors. In regime $\I$ where $\ma$ is typically very small (Fig.~\ref{fig:tseries_figure} 
A and D), there is one or more `mainstream' opinion groups near $x \sim 1/2$ with a majority of the agents in the system, and a number of 
smaller, `extremist' opinion groups at positions closer to the boundaries $x = 0, 1$. The medium opinion stays on average at the center of the 
opinion space, close to the mainstream group(s), and although continuously contested by editors with extremist views, it remains stable and leads 
to a very slow convergence towards consensus. The reason for a long relaxation time in regime $\I$ is intuitively clear: for low $\ma$ any change 
in AMD is small and thus both medium and extremist opinions fail to converge quickly. When the tolerance $\et$ decreases the effect is even more striking; 
even though the number of opinion groups is larger (according to the approximation $n_c \sim 1/[2\et]$), the article is quite stable and 
remains close to the mainstream view.

In regime $\II$ identified with intermediate values of $\ma$ (Fig.~\ref{fig:tseries_figure} B and E), the fixed pool dynamics produces quasi-periodic 
oscillations in the medium value $A$, which appear after an initial stage of opinion group formation and end up very quickly in total consensus. 
Quite surprisingly, the final consensual opinion is not $x \sim 1/2$ (as in regime $\I$) or that of the initial mainstream group, but some 
intermediate value closer to the extremist groups at the boundaries. This is indicative of a symmetry-breaking transition: as $\ma$ increases, 
a symmetric stationary state at $x \sim 1/2$ is replaced by a final state close to 0 or 1. The oscillations in regime $\II$ can initially be 
understood as a struggle over
medium dominance among the different opinion groups created by AAD. The AMD mechanism couples the medium dynamics with these groups, exchanging 
agents between them and thus modifying their positions, until the majority group wins over the rest and consensus is achieved. For small $\et$ 
oscillations are more well-defined and last for longer, while extremist groups tend to diffuse towards the mainstream.

In regime $\III$ for large $\ma$ (Fig.~\ref{fig:tseries_figure} C and F), extremist agents directly converge to a mainstream group and an article 
at the center. Since in this case $\ma$ is so large, after any jump of the article extremist agents can enter its tolerance interval and start 
drifting inwards. The limiting condition for this behavior is $\ma=1-\ea/(1/2-\ea)$~\cite{torok2013opinions}, a line separating regime $\III$ from 
the rest. A smaller $\et$ value produces a more erratic medium evolution, with occasional jumps up and down.

\begin{figure}[ht!]
\includegraphics[width=0.95\linewidth]{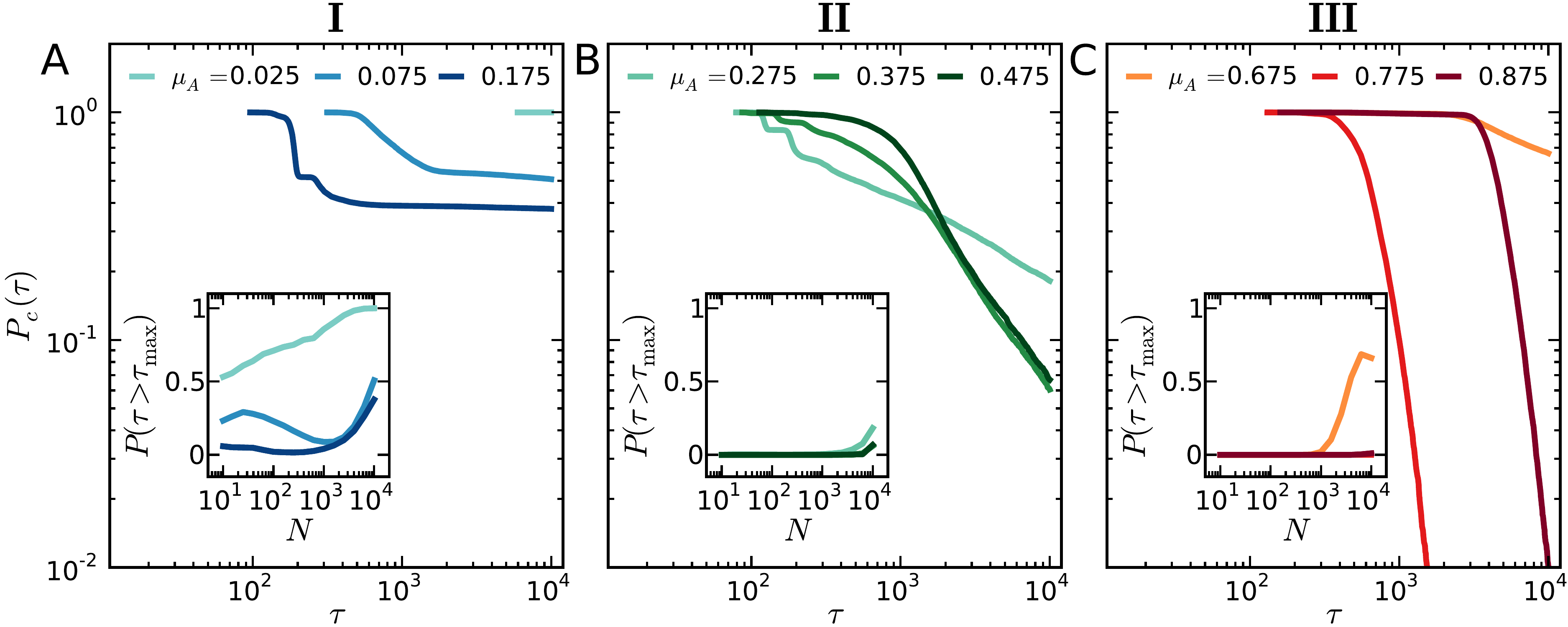}
\caption{\csentence{Distributions of relaxation time.} (A, B, C) Cumulative 
distribution $P_c(\tau)$ of the relaxation time $\tau$ necessary to reach consensus and thus end the dynamics, 
for different values of the medium convergence $\ma$. Insets: Probability $P(\tau > \tmax)$ that the relaxation 
time is larger than $\tmax = 10^4$ (the maximum time allowed in the numerical simulations), as a function of $N$ 
for selected values of $\ma$. The symbols $\I$, $\II$ and $\III$ denote the three 
different regimes found in the dynamics. Simulations correspond to $\et = 0.2$, $\ea = 0.1$ and $N = 10^4$, 
with averages over $10^4$ realizations.}
\label{fig:scaling_figure}
\end{figure}

The regimes of the fixed agent pool dynamics can be quantified on average by taking a look at the cumulative distribution $P_c(\tau)$ of the 
relaxation time $\tau$ (Fig.~\ref{fig:scaling_figure}). In regime $\I$ the tail of $P_c(\tau)$ is quite flat, getting flatter as $\ma$ decreases. 
In contrast, the distribution has a power-law and an exponential tail in regimes $\II$ and $\III$, respectively, signaling shorter relaxation 
times. The only exception is the transition between $\II$ and $\III$, where $\tau$ might be as large as in $\I$. Since $P_c(\tau)$ tends to be 
broad, the average value of $\tau$ is not very meaningful and we opt instead for the probability $P(\tau > \tmax)$ that the relaxation time is 
larger than a fixed maximum time. Numerically, we estimate $P(\tau > \tmax)$ as the fraction of realizations of the dynamics that have not reached 
consensus after $\tmax$ time steps, 
out of a large total of $10^4$ realizations. In regimes $\II$ and $\III$, $P(\tau > \tmax)$ remains small as $N$ increases, indicating that $\tau$ is roughly 
independent of system size. On the other hand, $P(\tau > \tmax)$ scales with $N$ for $\I$ and for the boundary between $\II$ and $\III$, even reaching 1 for 
appropriate values of $\ma$ and $N$. A corollary is that even modestly-sized systems may only reach consensus after an astronomical time, if 
the medium convergence value is appropriate.

\begin{figure}[ht!]
\includegraphics[width=0.95\linewidth]{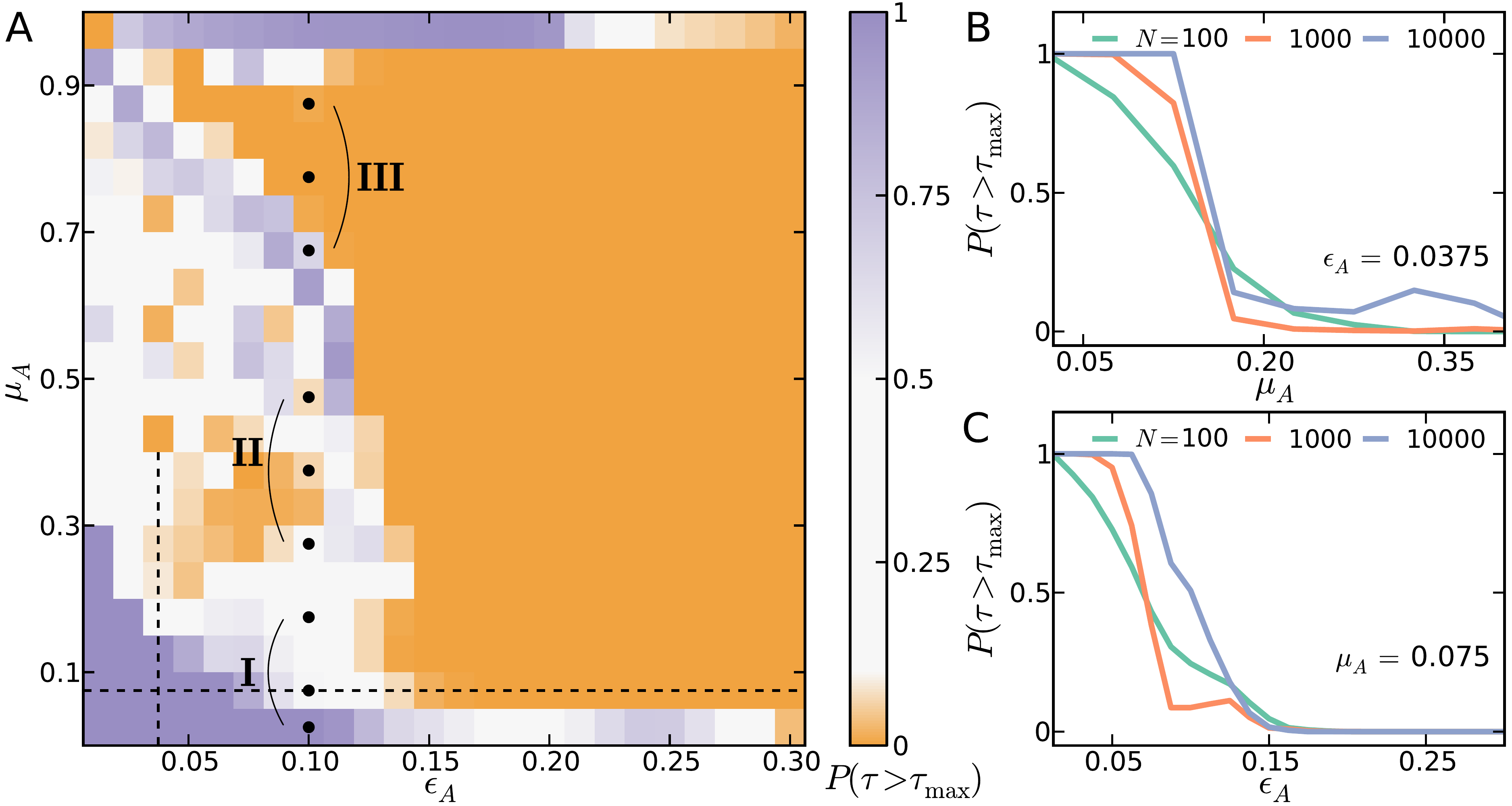}
\caption{\csentence{Phase diagram for a fixed agent pool.} (A) Phase diagram in $(\ea, \ma)$ space of the 
probability $P(\tau > \tmax)$ that the relaxation time is larger than $\tmax = 10^4$, in a system of size $N = 10^4$. Points 
give the $(\ea, \ma)$ values used in Fig.~\ref{fig:scaling_figure}, 
corresponding to regimes $\I$, $\II$ and $\III$. (B, C) Cross sections of the phase diagram along the dashed lines in (A) for 
varying $N$. Simulations correspond to $\et = 0.2$, with averages over $10^4$ realizations.}
\label{fig:phasediag_figure}
\end{figure}

The transition between regimes becomes even clearer when we consider the effect of the medium tolerance $\ea$, resulting in a phase 
diagram for $P(\tau > \tmax)$ in $(\ea, \ma)$ space (Fig.~\ref{fig:phasediag_figure} A). It turns out that regimes $\I$ and $\II$ cover most 
of the low $\ea$ values, while the line $\ma=1-\ea/(1/2-\ea)$ roughly signals the transition to regime $\III$, which covers a broad area of 
large $\ea$. As $N$ increases, the transition to $\I$ from either $\II$ or $\III$ (Fig.~\ref{fig:phasediag_figure} B and C) becomes sharper: 
a consensual final state reached after a very short time gives way to a stationary state that remains stable for really long times. 
Such features of the phase diagram remain qualitatively unchanged if we substitute $P(\tau > \tmax)$ with another measure giving 
robust statistics, such as the median relaxation time of the dynamics.

\begin{figure}[ht!]
\includegraphics[width=0.95\linewidth]{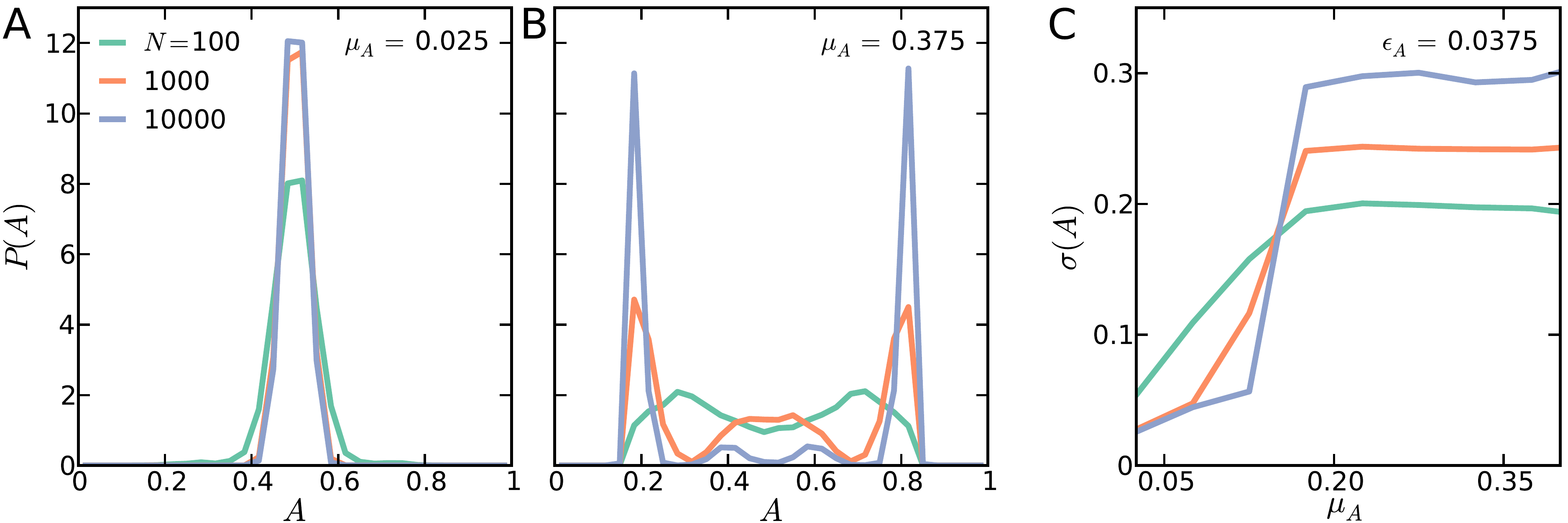}
\caption{\csentence{ Symmetry-breaking transition.} (A, B) Distribution $P(A)$ of the final medium value $A$ reached after a 
time $\min(\tau, \tmax)$ has elapsed, for varying $N$. The selected $\ma$ values represent regimes $\I$ (A) and $\II$ (B). (C) 
Standard deviation $\sigma(A)$ of the final medium value as a function of $\ma$, for several values of $N$. This order parameter 
signals a symmetry-breaking transition between regimes $\I$ and $\II$. Simulations correspond 
to $\et = 0.2$, $\ea = 0.0375$ and $\tmax = 10^4$, with averages over $10^4$ realizations.}
\label{fig:symmetry_figure}
\end{figure}

Finally, we can consider the symmetry-breaking transition between regimes $\I$ and $\II$ by taking a look at the density distribution $P(A)$ 
of the final medium value (Fig.~\ref{fig:symmetry_figure} A and B). After either $\tau$ or $\tmax$ has passed, the majority of opinions are 
in consensus with $A$, making $P(A)$ a good approximation for the final opinion distribution $P(x)$ as well. In regime $\I$ the medium 
distribution is roughly unimodal and peaked at $A \sim 1/2$, signaling a stable and moderate medium. Here the relaxation time is quite long and for most realizations $\tau > \tmax$. 
In regime $\II$, however, $P(A)$ becomes 
bimodal, meaning that the medium is more likely to end up close to the extremes rather than in the center. When $N$ is large, the main peaks in $P(A)$ correspond to consensual final states with $\tau \leq \tmax$, while the secondary peaks are made up of long-lived realizations with long relaxation time. Larger values of $\tmax$, although computationally expensive, would therefore let us see a strictly bimodal medium distribution for regime $\II$. 
As $N$ increases the distribution peaks become sharper and we 
can use the standard deviation $\sigma (A)$ of the final medium value as an order parameter for the transition (Fig.~\ref{fig:symmetry_figure} C). 
In the thermodynamic limit $N \to \infty$, a vanishing $\sigma (A)$ for $\I$ implies a stationary stable state with $A \sim 1/2$ and no consensus. 
As $\ma$ increases this symmetry gets broken, $\sigma (A)$ becomes nonzero and a true final state of consensus appears.

This symmetry-breaking mechanism may be understood analytically via a rate equation formalism~\cite{torok2013opinions}. The resulting rate equation 
can be solved numerically assuming three editor groups: a mainstream at $x \sim 1/2$ and two extremists with opinions close to the boundaries. The 
solution shows stability for the medium at the mainstream opinion when $\ma$ is small, but becomes unstable and oscillating for $\ma \simeq 3\ea\pm0.1$. 
The bifurcation transition is very sensitive on the position of the extremists, depending not only on $(\ma,\ea)$ but also on the initial conditions. 
This is in part the cause of the `noisy' landscape of regime $\II$ in Fig.~\ref{fig:phasediag_figure} A,  which appears regardless of the measure used to draw the phase diagram.

\subsection{Agent renewal}
\label{ssec:openPool}

In real systems the pool of collaborators is usually not fixed: Editors come
and go and very often the number of editors fluctuates in time as
external events may incite more or less attention. To keep things
simple we only focus on systems with a fixed number of editors ($N$
agents), but we allow agent replacement with probability $\pn \neq 0$. In our numerical simulations this
happens prior to editing, and new agents have initially random opinions coming from a uniform distribution.

If $\ea < 1/2$ there is always an opinion range outside the article
tolerance region $[A-\ea, A+\ea]$ and new agents may enter such range and edit the article. From WP data we know
that even peaceful articles have few disputes now and then so such a
scenario is realistic. This is thus in contrast with the case of a fixed
opinion pool, where consensus is theoretically always achieved.

A stronger statement can be shown \cite{torok2013opinions},
namely that if
\begin{equation}
\ea>\eac=\frac{1}{2-\ma}
\end{equation}
then consensus is always reached after a finite number of steps, but if
$\ea < \eac$ there are realizations that do not reach consensus ever.
We show here an example: if the medium value is $A = \eac$, then for
$\ea = \eac-\varepsilon$ an editor at $x=0$ will disagree with
the article and change it by $\Delta = \eac\ma$, so the new medium value would be $A = 1-\eac$. Afterwards an agent at $x=1$ can 
restore the article to its previous state and avoid consensus.

The lack of full consensus does not mean that the system is always in
a conflict state. There are periods when $A$ remains
unchanged and these peaceful times are ended by conflicts in which the
opinion of the article is continuously disputed between agent groups
of different opinion. If the dispute is settled (i.e. all agents are
satisfied by the article) a new peaceful period may start. The ratio of
these peaceful and conflicting periods changes with the parameters and
may be considered as a good candidate for an order parameter. Thus we
define the order parameter $P$ as the relative length of the peaceful
periods.

\begin{figure}[ht!]
\includegraphics[width = .7 \linewidth]{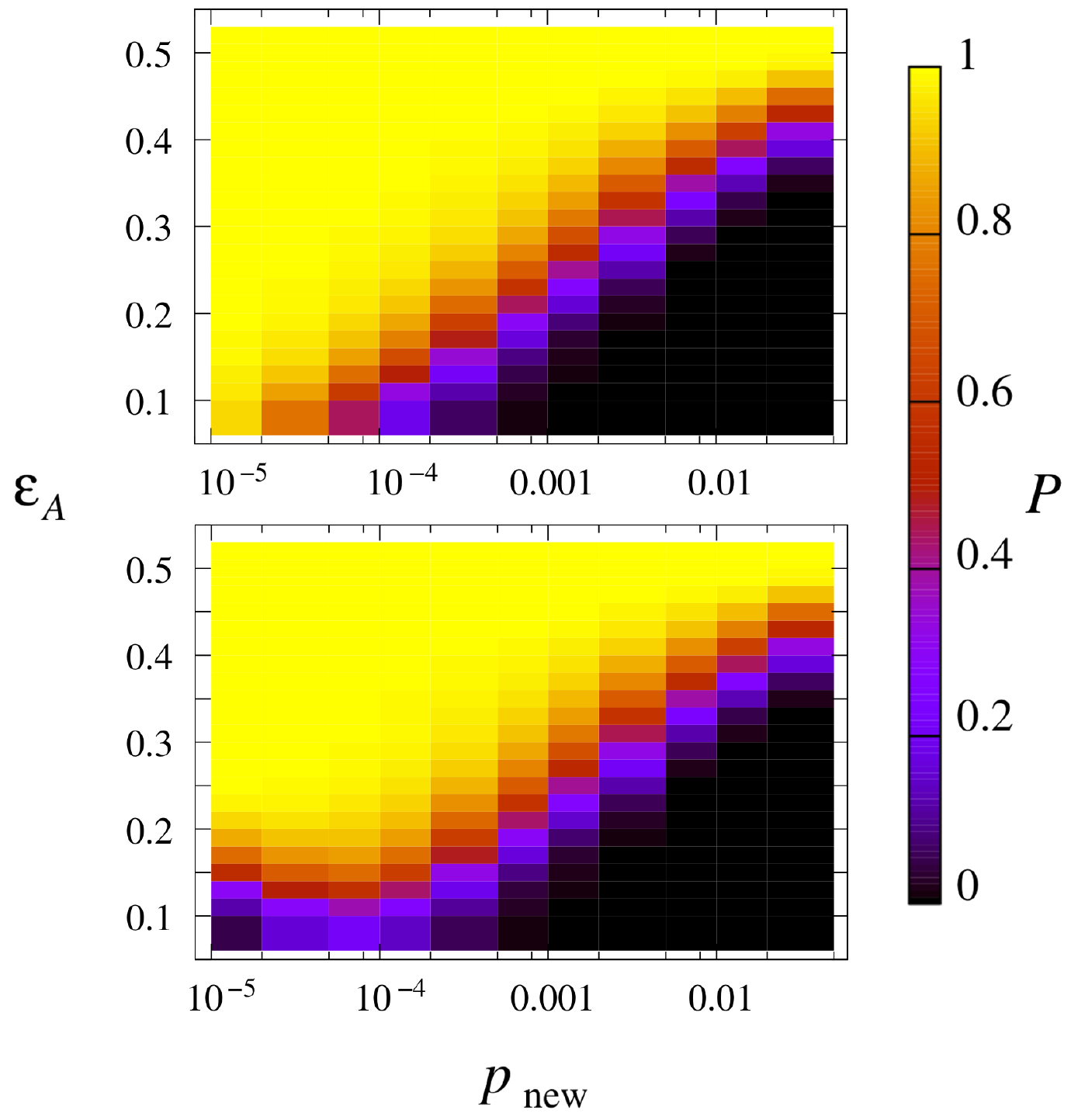}
\caption{\label{Fig:3d_phase}\csentence{Order parameter for the agent
renewal case.} Order parameter $P$ as a function of the medium tolerance $\ea$ and
the agent replacement rate $\pn$ for systems of size $N = 80$. The
chosen initial condition is consensus for the top diagram and conflict for the
bottom one.}
\end{figure}

The order parameter is plotted for two different initial conditions in
Fig.~\ref{Fig:3d_phase}. The top figure shows the value of the order
parameter $P$ for a `peaceful' initial condition when all agents had the opinion
$x_i = 1/2$. The bottom figure was instead obtained for a system with `conflict'
initial conditions, namely one with 20\% of agents divided between two
extremist groups of opinions 0 and 1 (and the rest at $x_i = 1/2$)
before the start of the dynamics.

\begin{figure}[ht!]
\includegraphics[width = .7 \linewidth]{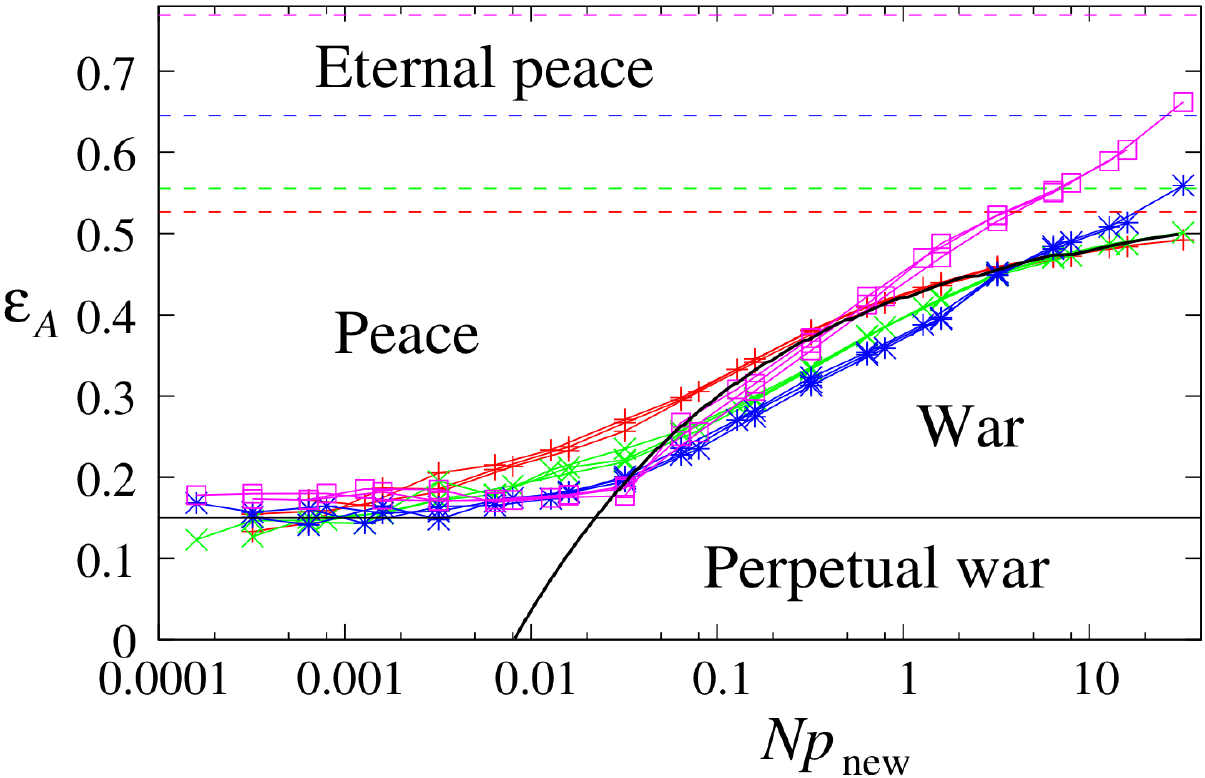}
\caption{\label{Fig:noisyphasediag}\csentence{Phase diagram for a system with agent renewal.} 
Largest gradient of $P$ by using the lower plot in
Fig.~\ref{Fig:3d_phase}, for varying $\ma$. Simulations correspond to
$\et=0.2$ and system size ranging from $N = 10$ to $640$. The article
convergence parameter was $\ma=0.1$, $0.2$, $0.45$, $0.7$ for red,
green, blue and magenta respectively. The curved black line is the
analytical result for $\ma=0.1$. The horizontal line limiting the
prepetual peace domain is at $\ea=0.15$. The eternal peace is limited
by $\eac$ (shown with dashed lines for the same color) which depends
on $\ma$.}
\end{figure}

It is clear that there are two distinct regimes in the phase diagram
of Fig.~\ref{Fig:3d_phase}: one characterized by $P=1$ (`peaceful' regime), the other with $P=0$ (`conflict' regime) and a
sharp transition in between. There is a region which is different in
the two cases and will be discussed later. We then identify the transition
point with the largest gradient of $P$ by using the lower plot in
Fig.~\ref{Fig:3d_phase}. The resulting phase diagram is shown
in Fig.~\ref{Fig:noisyphasediag}.

\begin{figure}[ht!]
\includegraphics[width=.95\linewidth]{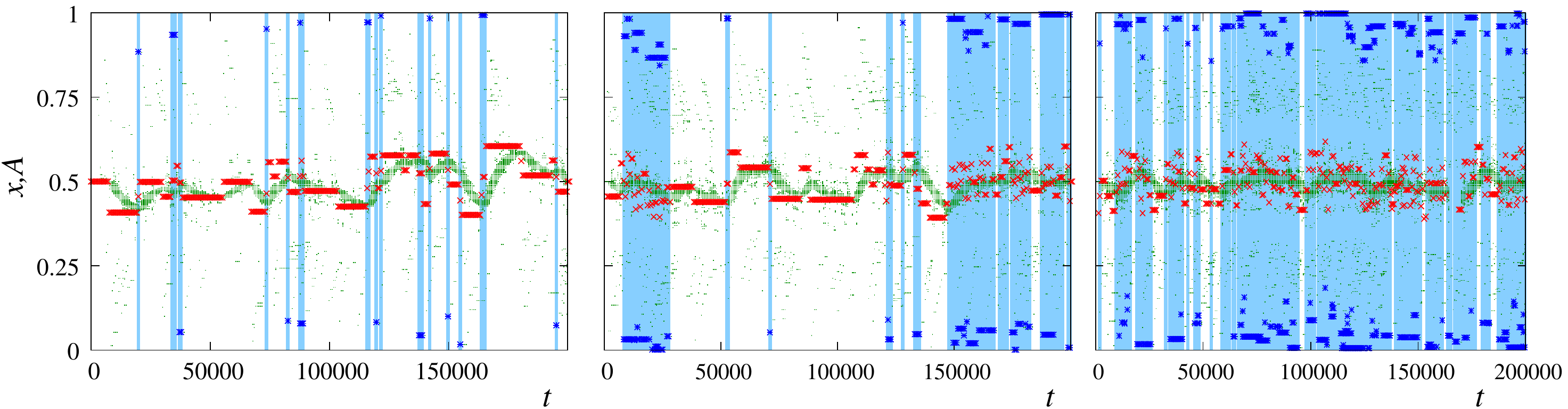}
\caption{\label{Fig:noisyevolsample}\csentence{Time evolution of
opinions.}
Samples of medium/agent opinions as a function of time for $\ea=0.42$,
and for three different regimes represented by $\pn=0.001, 0.002,
0.003$ (from left to right respectively). Colour coding is as follows:
Red points (opinion of the article), green dots (agents who are
satisfied with the article), blue points (agents whose opinion is
outside the medium tolerance interval), and light blue background
(conflict regions).}
\end{figure}

This transition is further illustrated in
Fig.~\ref{Fig:noisyevolsample} where we display sample time evolutions of the opinions
of agents and medium. The left panel shows an example
of a peaceful regime. As mentioned before, from time to
time new agents arrive with incompatible views with respect to the
article but they are pacified very fast, i.e. the conflict periods are
short. In the transition regime (middle panel) the scenario of
peaceful times interrupted by short conflicts is still observable,
but periods of continuous conflict occasionally appear. In the conflict regime exemplified by the right panel, these conflict bursts become
persistent and the peaceful periods tend to disappear.

The above transition is the result of a competition between two timescales. New
agents arrive outside of the article's tolerance interval with an `insertion'
timescale $\tau_\mathrm{ins}\propto N\pn$. In order to have $P > 0$
the conflicts must be resolved before a new extremist agent arrives.
Let us note that the convergence is very fast if there is only one
extremist group. The problem is solved by displacing the article
opinion by the required amount, which can be done in few ($N$
independent) steps.  This is what happens in the left panel of
Fig.~\ref{Fig:noisyevolsample}.  On the other hand, if we have two
extremist groups on both sides of the opinion interval the relaxation
is much slower and this is manifested in a much longer relaxation time.
Thus, at the transition the insertion timescale is equal to the
relaxation time of the case with two extremist groups, which is analogous to the
fixed agent pool version of the model.

The task here is to determine the relaxation time of the fixed pool
version of the model and relate it to $\tau_\mathrm{ins}$.
For large values of the medium tolerance ($\ea > 1/4$), the relaxation
time can by calculated analytically \cite{torok2013opinions},
\begin{equation}
\tau(e)= c(\ma)N\left([2 e^2+e_0^2  (n-1)] n - e e_0 (n-1) (2 + n)\right),
\end{equation}
where $e = \eac-\ea$, $e_0 = \eac-1/2$, $n$ denotes the integer
part of $e/e_0$ (which is actually the number of steps the medium can
make in one direction) and $c$ is a constant depending on $\ma$.

The above approach works well for $\ea>0.3$ and $\ma<0.3$ (regime
$\III$ of the fixed pool case). If the mainstream group gets dissatisfied
either by the large jump ($\ma$ is too large) or by the small
tolerance ($\ea$ too small) of the article, the reasoning presented in
\cite{torok2013opinions} breaks down and new effect comes into play,
namely the relaxation times of the fixed pool system becomes be
enormous (regime $\I$).

As we enter regime $\I$ of the fixed pool dynamics the relaxation time
increases sharply (see Fig.~\ref{fig:phasediag_figure} B and C). This
means that if the system gets into a conflict state it will remain
there for ever, which happens for,
\begin{equation}
\epsilon_{A,\mathrm{lim}}=\frac{1}{4}-\frac{\et}{2}.
\end{equation}

This is why, starting from a conflict initial condition, the lower phase diagram in
Fig.~\ref{Fig:3d_phase} shows $P = 0$ for $\ea<0.15$. On the other
hand, in order to initiate such a conflict one needs to have a
situation where two extremists appear on both ends of the opinion space
outside of the article tolerance interval.  If we have a single extremist
then the consensus will be reached within a few time steps,
independently of $N$. So the probability that we create a long-lasting
conflict state decreases proportionally to the agent replacement
probability. This is why we see only peace on the finite-time realizations
leading to the upper phase diagram in Fig.~\ref{Fig:3d_phase}. Had we waited long enough, a
conflict would have been formed for $\ea<1/4-\et/2$ and would have persisted
further on.

In summary, the typical behavior of our model in the presence of agent renewal may be divided into four distinct regimes:
\begin{enum}
\item {\em Eternal peace} ($\ea>\eac$): The system reaches
consensus very fast and remains there for ever.
\item {\em Peace} ($\ea>\frac{1}{4}-\frac{\et}{2}$ and above the phase transition
line): The system is mainly in a consensual state and only
interrupted by short disputes.
\item {\em War} ($\ea>\frac{1}{4}-\frac{\et}{2}$ and below the phase transition 
line): The system is mainly in a state of disagreement.
\item {\em Perpetual war} ($\ea<\frac{1}{4}-\frac{\et}{2}$): In this
regime and in the thermodynamic limit $N \to \infty$ no consensus may exist.
\end{enum}

\subsection{The case of Wikipedia}
\label{ssec:wikipedia}

Although the model described and analyzed above is simplified enough to be extendable to various 
cases of collaboration, we specially intend to use it to explain some of the empirical
observations regarding edit wars in WP. 

Wikipedia is huge, not only in its number of articles and users but in the number of times articles are 
edited. In most cases articles are not written in a collaborative way, i.e., they have single 
authors or a few authors who have written and edited different parts of the article without any significant 
interaction~\cite{kimmons2011understanding}. In contrast, a few cases show significant 
constructive and/or destructive interactions between editors. The latter situation has been named `edit war' 
by the WP community and defined as follows: ``An edit war occurs when editors who disagree about the content 
of a page repeatedly override each other's contributions, rather than trying to resolve the 
disagreement by discussion''~\cite{wiki:war}.

To start an empirical analysis of such opinion clashes and the way they are entangled with collaboration, we need to 
be able to locate and quantify edit wars.

\subsubsection{Controversy measure}

An algorithm to quantify edit wars and measure the amount of social clashes for WP articles 
has been introduced and validated before~\cite{sumi2011edit}, and then used to study extensively the 
dynamical aspects of WP conflicts~\cite{yasseri2012dynamics}. An independent study~\cite{sepehri2012leveraging} 
has also shown that this measure is among the most reliable in capturing very controversial articles. 

We quantify the `controversiality' of an article based on its edit history by focusing on `reverts' (i.e. 
when an editor completely undoes another editor’s edit and brings the article back to the state just before the last version).
Reverts are detected by comparing all pairs of revisions of an article 
throughout its history, namely by comparing the MD5 hash code~\cite{rivest1992md5} of the revisions. Specifically, a revert is 
detected when two versions in the history line are exactly the same. In this case the latest edit (leading 
to the second identical revision) is marked as a revert, and a pair of editors, referred to as reverting and 
reverted editors, are recognized. 

Very soon in our investigation we noticed that reverts can have different reasons and not in all cases 
signalize a conflict of opinions. For example, an editor could revert personal edit mistakes or someone else's. 
Reverts are also heavily used to suppress vandalism, in itself a different type of destructive social behavior, 
but with no collaborative intention and therefore out of our interest. Thus we narrowed down our analysis 
to `mutual reverts'. A mutual revert is recognized if a pair of editors $(x, y)$ is observed once with $x$ as the reverter and once with $y$.
We also noticed that mutual reverts between pairs of editors at different levels 
of expertise and experience in WP editing could contribute differently to an edit war. Two 
experienced editors getting involved in a series of mutual reverts is usually a sign of a more serious conflict, as
opposed to the case when two newbies or a senior editor and a newbie bite each other \cite{halfaker2011bite}. 
As a solution we introduced a `weight' for each editor, and to sum up all reverts within the history of an article we counted each revert 
by using the smaller weight of the pair of editors involved in it. The weight of an editor $x$ is defined as the 
number of edits performed by him or her, and the weight of a mutually reverting pair is defined as the 
minimum of the weights of the two editors. The controversiality $M$ of an article is then defined by summing 
the weights of all mutually reverting editor pairs, multiplying this number by the total number of editors $E$ 
involved in the article. Overall,
\begin{equation}\label{Eq:M}
 M = E \sum_{\text{all mutual reverts}} min(N^{\rm d}, N^{\rm r}),
\end{equation}
where $N^{\rm r}$, $N^{\rm d}$ are the number of edits for the article committed by the reverting/reverted editor. This 
measure can be easily calculated for each article, irrespective of the language, size, and length of its history.

Before starting our discussion about the empirical dynamics of conflict and its comparison with theoretical results, a remark on 
the most controversial articles in WP. We have calculated $M$ for all articles in 13 different languages, from the start of each language WP up to March 2010. In Table~\ref{top-10} we show the list of the top-10 most controversial articles. A more complete and detailed analysis of the lists of the most controversial WP articles in different languages and differences and similarities between them can be found elsewhere~\cite{yasseri2014controversial}.

\begin{table}[ht!]
\caption{List of the most controversial articles in different language WPs according to $M$.}
\begin{tiny}
\begin{tabular}{p{.0cm}p{2.2cm}p{2.2cm}p{2.2cm}p{2.2cm}p{2.2cm}}
\hline
&		\bf{English}		&		\bf{German}		&		\bf{French}		&		\bf{Spanish}		&		\bf{Portuguese}	\\
\hline
1		&		George W. Bush		&		Croatia		&		S\'gol\`ene Royal		&		Chile		&		S\~ao Paulo\\	
2		&		Anarchism		&		Scientology		&		Unidentified flying object		&		Club América		&		Brazil	\\
3		&		Muhammad		&		9/11 conspiracy theories		&		Jehovah’s Witnesses		&		Opus Dei		&		Rede Record	\\
4		&		List of WWE personnel		&		Fraternities		&		Jesus		&		Athletic Bilbao		&		José Serra	\\
5		&		Global warming		&		Homeopathy		&		Sigmund Freud		&		Andrés Manuel López Obrador		&		Grêmio Foot-Ball Porto Alegrense	\\
6		&		Circumcision		&		Adolf Hitler		&		September 11 attacks		&		Newell’s Old Boys		&		Sport Club Corinthians Paulista\\	
7		&		United States		&		Jesus		&		Muhammad al-Durrah incident		&		FC Barcelona		&		Cyndi Lauper	\\
8		&		Jesus		&		Hugo Chávez		&		Islamophobia		&		Homeopathy		&		Dilma Rousseff	\\
9		&		Race and intelligence		&		Minimum wage		&		God in Christianity		&		Augusto Pinochet		&		Luiz Inácio Lula da Silva	\\
10		&		Christianity		&		Rudolf Steiner		&		Nuclear power debate		&		Alianza Lima		&		Guns N’ Roses	\\
\hline
&		\bf{Czech}		&		\bf{Hungarian}		&		\bf{Romanian}		&		&	\\
\hline
1	&		Homosexuality		&		Gypsy Crime		&		FC Universitatea Craiova		&		&	\\
2	&		Psychotronics		&		Atheism		&		Mircea Badea		&		&	\\
3	&		Telepathy		&		Hungarian radical right		&		Disney Channel (Romania)		&		&	\\
4	&		Communism		&		Viktor Orb\'an		&		Legionnaires' rebellion \& ucharest pogro	&		&	\\
5	&		Homophobia		&		Hungarian Guard Movement		&		Lugoj		&		&	\\
6	&		Jesus		&		Ferenc Gyurcsány’s speech in May 2006		&		Vladimir Tismăneanu		&		&	\\
7	&		Moravia		&		The Mortimer case		&		Craiova		&		&	\\
8	&		Sexual orientation change efforts		&		Hungarian Far- right		&		Romania		&		&	\\
9	&		Ross Hedvíček		&		Jobbik		&		Traian Băsescu		&		&	\\
10	&		Israel		&		Polgár Tamás		&		Romanian Orthodox Church		&		&	\\
\hline	
	&		\bf{Arabic}		&		\bf{Persian}	&		\bf{Hebrew}	&		\bf{Japanese}	&		\bf{Chinese}	\\
\hline
1	&		Ash’ari		&		Báb		&		Chabad		&		Koreans in Japan		&		Taiwan	\\
2	&		Ali bin Talal al Jahani		&		Fatimah		&		Chabad messianism		&		Korea origin theory		&		List of upcoming TVB series	\\
3	&		Muhammad		&		Mahmoud Ahmadinejad		&		2006 Lebanon War		&		Men's rights		&		TVB	\\
4	&		Ali 		&		People's Mujahedin of Iran		&		B'Tselem		&		internet right-wing		&		China	\\
5	&		Egypt		&		Criticism of the Quran		&		Benjamin Netanyahu		&		AKB48		&		Chiang Kai-shek	\\
6	&		Syria		&		Tabriz		&		Jewish settlement in Hebron		&		Kamen Rider Series		&		Ma Ying-jeou	\\
7	&		Sunni Islam		&		Ali Khamenei		&		Daphni Leef		&		One Piece		&		Chen Shui-bian	\\
8	&		Wahhabi		&		Ruhollah Khomeini		&		Gaza War		&		Kim Yu-Na		&		Mao Zedong	\\
9	&		Yasser Al-Habib		&		Massoud Rajavi		&		Beitar Jerusalem F.C.		&		Mizuho Fukushima		&		Second Sino-Japanese War	\\
10	&		Arab people		&		Muhammad		&		Ariel Sharon		&		GoGo Sentai Boukenger		&		Tiananmen Square protests of 1989	\\
\end{tabular}
\end{tiny}  \label{top-10}
\end{table}

\subsubsection{Dynamics of conflict and war scenarios}

\begin{figure}[ht!]
\includegraphics[width=0.95\linewidth]{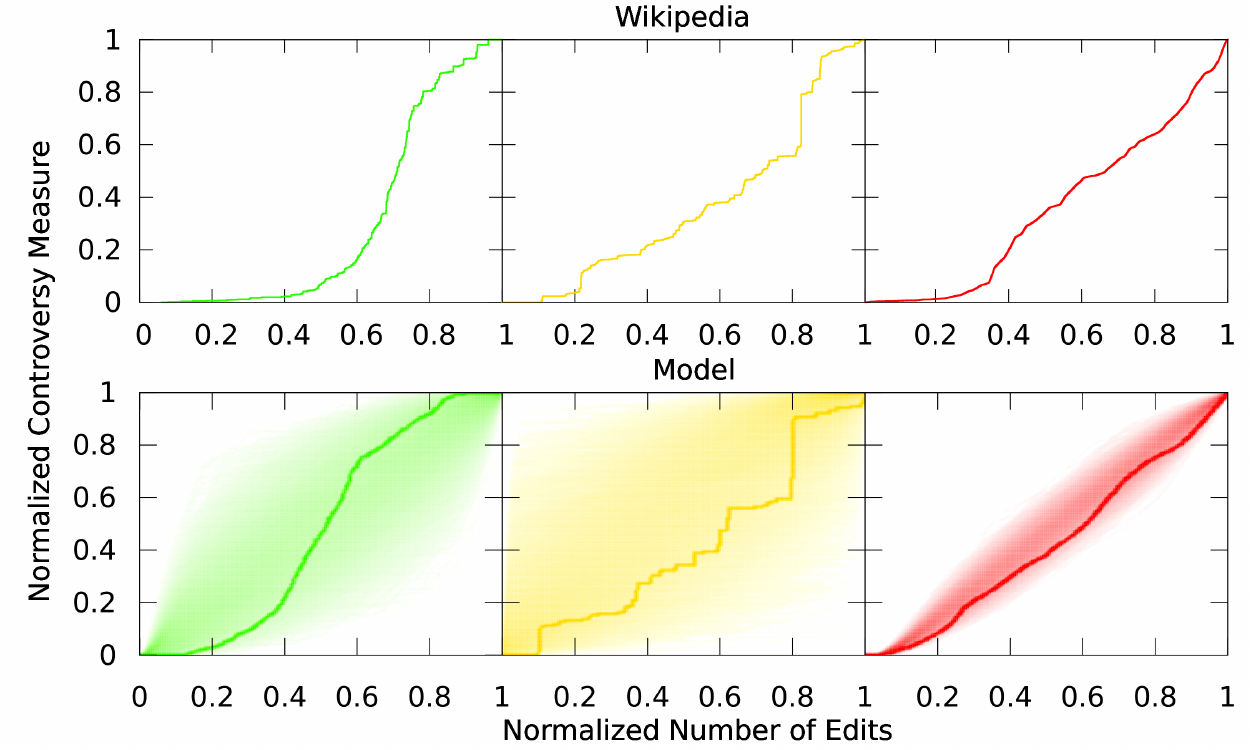}
\caption{\label{Fig:war-scenario} \csentence{War scenarios for WP and our model.} Top: Empirical controversy measure $M$ as a 
function of the number of article edits in three different war scenarios. From left to right, sample articles 
are `Jyllands-Posten Muhammad cartoons controversy', `Iran', and `Barack Obama'; and correspond to the 
regimes of `single war', `war-peace cycles', and `never-ending war' respectively. Bottom: Theoretical conflict 
measure $S$ in the case of agent renewal, reproducing the qualitatively analogue evolution 
of WP articles with parameter values $N=640$, $\et=0.2$ and $\ma=0.1$, as well 
as $\ea=0.35,  0.42, 0.30$ and $\pn= 0.001, 0.001, 0.002$ for the three war scenarios, 
respectively. Continuous lines correspond to selected single runs of the model, while the shading indicates the density of $S$ over an ensemble of $10^4$ realizations.}
\end{figure}

Measuring $M$ can not only lead us to rank the articles based on their cumulative controversy measure, 
but also enables us to follow edit wars in time as they emerge and get resolved, by investigating 
the evolution of $M$ as time passes and the article develops. In the top row of
Fig.~\ref{Fig:war-scenario} we show the time evolution 
of $M$ for three different sample articles. 

Based on the way $M$ evolves in time, we may categorize almost all controversial 
articles into three categories:
\begin{enum}
\item {\it Single war to consensus}: In most cases controversial articles can be included in this category. A single 
edit war emerges and reaches consensus after a while, stabilizing quickly. If the topic of the article is not particularly dynamic, the 
reached consensus holds for a long period of time (top left in Fig.~\ref{Fig:war-scenario}). 
\item {\it Multiple war-peace cycles}: In cases where the topic of the article is dynamic but the rate of new events (or production 
of new information) is not higher than the pace to reach consensus, multiple cycles of war and peace may
appear (top center in Fig.~\ref{Fig:war-scenario}).
\item {\it Never-ending wars}: Finally, when the topic of the article is greatly contested in the real 
world and there is a constant stream of new events associated with the subject, the article tends not to reach a 
consensus and $M$ increases monotonically and without interruption (top right in Fig.~\ref{Fig:war-scenario}). 
\end{enum}

The empirical war scenarios described previously are in qualitative agreement with the theoretical regimes of our model in the case of 
agent renewal, as seen from both the sample time series in Fig.~\ref{Fig:noisyevolsample} and the regimes of war and peace in the phase 
diagrams of Fig.~\ref{Fig:3d_phase} and Fig.~\ref{Fig:noisyphasediag}. Unfortunately, the theoretical order parameter $P$ is quite difficult to
measure in real systems as editor opinions are not known. What we know
instead is the controversy measure $M$ of Eq.~\ref{Eq:M}. As mentioned before, $M$
counts conflict events (i.e. mutual reverts) and weights them by the
maturity of the editor. This process can actually be repeated for the model:
The editor maturity $T_i$ is then defined as the number of time steps an agent has been in
the pool of editors (a quantity constantly reset by agent replacement), and a
conflict event is considered as the time an editor modifies the article, since this implies the agent is not satisfied with the state of the medium.

\begin{figure}[ht!]
\includegraphics[width=0.8\linewidth]{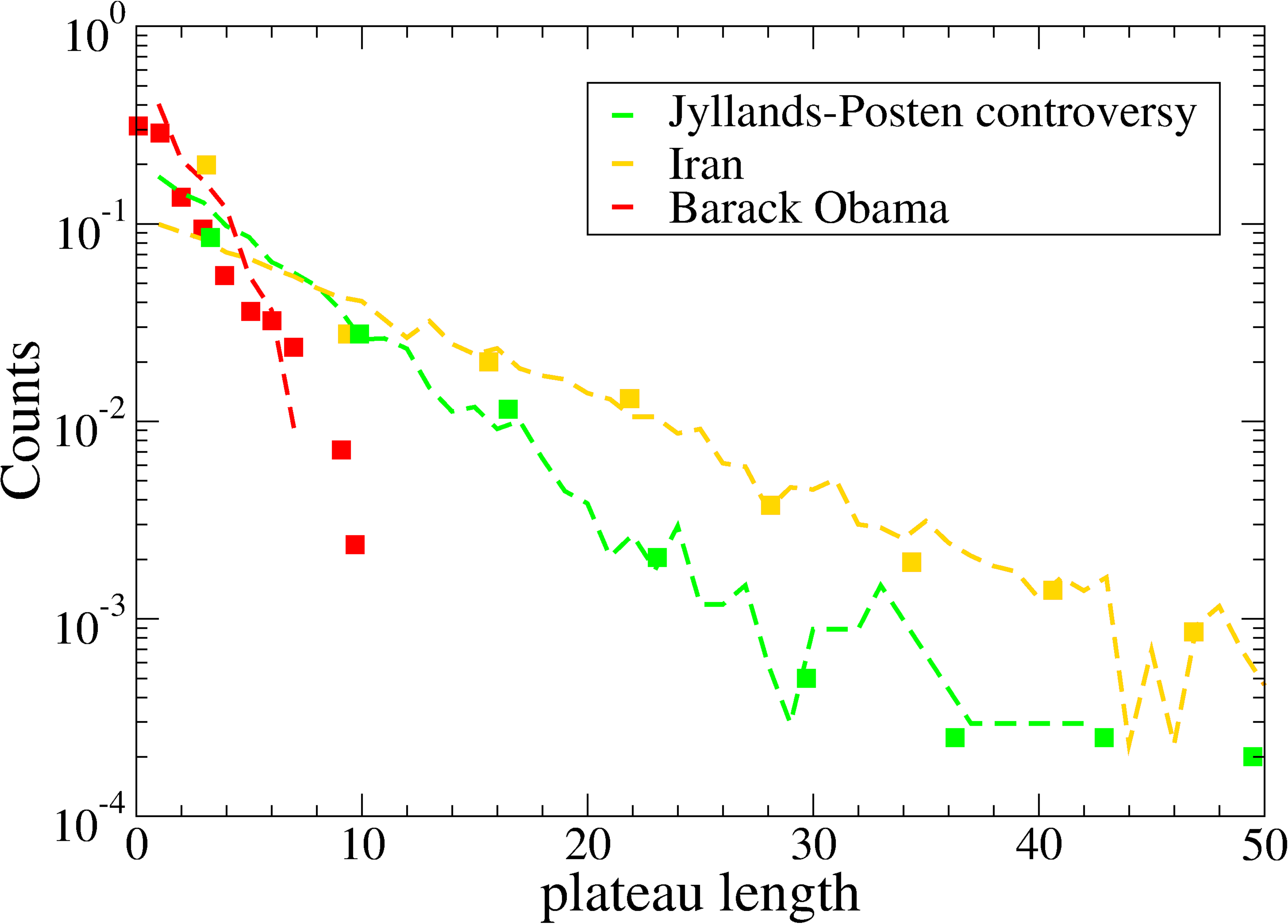}
\caption{\label{Fig:plateau} \csentence{Peace periods in WP and our model.} Distribution of plateau lengths for selected 
articles in WP (squares) and tuned parameters in our model (lines) for the three war scenarios shown in Fig.~\ref{Fig:war-scenario}. 
The length of a plateau or peace period is defined as the number of edits between two successive increments in either $M$ or $S$.}

\end{figure}

Thus a theoretical counterpart $S$ to the WP controversy measure $M$ may be defined as follows: Let $S=0$ at the beginning of the dynamics. Then in each update $t^*$ (out of the $N$ that constitute a time step in the dynamics), when editor $i$ changes the state of the article by the amount $\Delta = |A(t^*+1)-A(t^*)|$ we increment $S$ by $T_i\Delta$, where $T_i$ measures the time $i$ has been in the editorial pool. Examples of the temporal evolution of $S$ (lower row in
Fig.~\ref{Fig:war-scenario}) closely reproduce the qualitative
behavior of $M$ for different war scenarios. To further compare empirical observations in WP with our model predictions, we measure the 
typical length of a constant `plateau' in the $M$ and $S$ time series, i.e. the number of edits between two successive increments. As seen in the distribution of plateau length for WP and the model (Fig.~\ref{Fig:plateau}), a statistical agreement for all three war scenarios is clear.

A last word on WP banning statistics. A way of estimating the number of extremists is to count the number of editors who have been `banned' 
from editing. Explicitly, ``a ban is a formal prohibition from editing some or all WP pages, either temporarily 
or indefinitely''~\cite{wiki:ban}. Usually banning is used against vandals and/or editors who violate WP 
policies, especially those related to edit wars. In Table~\ref{table:ban} we give some statistics of editors at 
different classes of editing activity, according to their number of edits. Interestingly, the 
relative population of banned editors is larger among more experienced editors (i.e. editors with more than 1000 edits). 
In other words, up to almost 20\% of experienced editors could have been involved in edit wars.
This is in complete accord with the choices we have made for the modeling setup, namely having two active extremist groups with roughly 20\% of the total number of editors.

\begin{table}[ht!]
 \caption{Percentage of banned users to the total number of editors at three different classes.}
\begin{tiny}
 \begin{tabular}{p{.9cm}p{1.3cm}p{1.2cm}p{1cm}p{1.2cm}p{1.2cm}p{1cm}p{1.2cm}p{1.2cm}p{1cm}}
\hline
  &	Num. Editors w. \textgreater 1000 Edits	&	Ban. Editors w.  \textgreater 1000 Edits	&	\% ban. w. \textgreater 1000 edits	&	Num. Editors w. \textgreater 100 Edits	&	Ban. Editors w.  \textgreater 100 Edits	&	\% ban. w. \textgreater 100 edits & Num. Editors w. \textgreater 1 Edits	&	Ban. Editors w. \textgreater 1 Edits	&	\% ban. w. \textgreater 1 edits\\
\hline
English	&	36280	&	6114	&	0.17	&	189174	&	20342	&	0.11	&	4552685	&	403851	&	0.09	\\
German	&	8714	&	1561	&	0.18	&	34777	&	3458	&	0.10	&	511291	&	31996	&	0.06	\\
French	&	5286	&	694	&	0.13	&	21940	&	1700	&	0.08	&	394385	&	16681	&	0.04	\\
Spanish	&	3834	&	765	&	0.20	&	19135	&	2404	&	0.13	&	479305	&	21850	&	0.05	\\
Portuguese 	&	1733	&	345	&	0.20	&	8077	&	1015	&	0.13	&	194584	&	7486	&	0.04	\\
Czech	&	700	&	112	&	0.16	&	2439	&	236	&	0.10	&	48030	&	2663	&	0.06	\\
Hungarian	&	844	&	138	&	0.16	&	3107	&	276	&	0.09	&	49024	&	1201	&	0.02	\\
Romanian	&	437	&	53	&	0.12	&	1675	&	130	&	0.08	&	36631	&	914	&	0.02	\\
Arabic	&	610	&	96	&	0.16	&	2736	&	198	&	0.07	&	80498	&	1085	&	0.01	\\
Persian	&	580	&	151	&	0.26	&	2531	&	406	&	0.16	&	56805	&	2544	&	0.04	\\
Hebrew	&	1009	&	233	&	0.23	&	3898	&	515	&	0.13	&	53318	&	4341	&	0.08	\\
Japanese	&	4010	&	786	&	0.20	&	19090	&	2845	&	0.15	&	242621	&	21995	&	0.09	\\
Chinese	&	2106	&	378	&	0.18	&	9002	&	1072	&	0.12	&	160579	&	9387	&	0.06	\\
\end{tabular}
\end{tiny}
\label{table:ban}
\end{table}

\section{Discussion and Conclusion}
\label{sec:conc}

Here we have studied through modeling the emergence, persistence and resolution of conflicts in a collaborative environment of humans such as 
WP. The value production process takes place through interaction between peers (editors for WP) and through direct modification of 
the product or medium (an article). While in most cases this process is constructive and peaceful, from time to time severe conflicts emerge.
We modeled the dynamics of conflicts during collaboration by coupling opinion formation with article editing in a generalized bounded-confidence 
dynamics. The simple addition of a common value-production process leads to the replacement of frozen opinion groups (typical of the 
bounded-confidence dynamics)
with a global consensus and a tunable relaxation time. The model with a fixed pool shows a rich phase diagram with several characteristic 
behaviors: a) an extremely long relaxation time, b) fast relaxation preceded by oscillating behavior of the medium opinion, and c) an even 
faster relaxation with an erratic medium. We have observed a symmetry-breaking, bifurcation transition between regimes a) and b), as well as 
divergence of the relaxation time in the transition between regimes b) and c).

If the pool is not fixed and editors are exchanged with new ones at a given rate, we obtain two different phases:
conflict and peace. A conflict measure can be defined for the modeled system and be directly compared to its empirical counterpart in 
real WP data. It is then possible to follow the temporal evolution of this measure of controversy and obtain a good qualitative agreement 
with the empirical observations. These results lead us to plausible explanations for the spontaneous emergence of current WP policies, 
introduced to moderate or resolve conflicts. 

Two remarks are at place here. In this study we have used a particular collaboration environment and compared our results with WP. 
The main reason behind is that for the free encyclopedia we have a full documentation of actions; however, we should emphasize that as web-based 
collaborative environments are abundant, we believe that our approach and results are much more general. Second, we are aware of the fact that 
the model contains a number of stringent simplifications: There are cultural differences between the WPs (e.g., in the usage of the talk 
page), and as in all human-related features there are large inhomogeneities in the opinions, in the tolerance level and in the activity of 
editors. Some of these aspects are under current study and will be taken into account for future research.


\begin{backmatter}

\section*{Competing interests}
The authors declare that they have no competing interests.

\section*{Author's contributions}
All authors designed the research and participated in the writing of the manuscript. GI and JT 
performed the numerical calculations and analytical approximations. TY analyzed the empirical data.

\section*{Acknowledgements}
The authors acknowledge support from the ICTeCollective EU FP7 project. JK thanks FiDiPro (TEKES) 
and the DATASIM EU FP7 project for support. JT thanks the support of
European Union and the European Social Fund through project
FuturICT.hu (grant no.: TAMOP-4.2.2.C-11/1/KONV-2012-0013).

\bibliographystyle{bmc-mathphys} 
\bibliography{literature} 

\begin{thebibliography}{54}
\ifx \bisbn   \undefined \def \bisbn  #1{ISBN #1}\fi
\ifx \binits  \undefined \def \binits#1{#1}\fi
\ifx \bauthor  \undefined \def \bauthor#1{#1}\fi
\ifx \batitle  \undefined \def \batitle#1{#1}\fi
\ifx \bjtitle  \undefined \def \bjtitle#1{#1}\fi
\ifx \bvolume  \undefined \def \bvolume#1{\textbf{#1}}\fi
\ifx \byear  \undefined \def \byear#1{#1}\fi
\ifx \bissue  \undefined \def \bissue#1{#1}\fi
\ifx \bfpage  \undefined \def \bfpage#1{#1}\fi
\ifx \blpage  \undefined \def \blpage #1{#1}\fi
\ifx \burl  \undefined \def \burl#1{\textsf{#1}}\fi
\ifx \doiurl  \undefined \def \doiurl#1{\textsf{#1}}\fi
\ifx \betal  \undefined \def \betal{\textit{et al.}}\fi
\ifx \binstitute  \undefined \def \binstitute#1{#1}\fi
\ifx \binstitutionaled  \undefined \def \binstitutionaled#1{#1}\fi
\ifx \bctitle  \undefined \def \bctitle#1{#1}\fi
\ifx \beditor  \undefined \def \beditor#1{#1}\fi
\ifx \bpublisher  \undefined \def \bpublisher#1{#1}\fi
\ifx \bbtitle  \undefined \def \bbtitle#1{#1}\fi
\ifx \bedition  \undefined \def \bedition#1{#1}\fi
\ifx \bseriesno  \undefined \def \bseriesno#1{#1}\fi
\ifx \blocation  \undefined \def \blocation#1{#1}\fi
\ifx \bsertitle  \undefined \def \bsertitle#1{#1}\fi
\ifx \bsnm \undefined \def \bsnm#1{#1}\fi
\ifx \bsuffix \undefined \def \bsuffix#1{#1}\fi
\ifx \bparticle \undefined \def \bparticle#1{#1}\fi
\ifx \barticle \undefined \def \barticle#1{#1}\fi
\ifx \bconfdate \undefined \def \bconfdate #1{#1}\fi
\ifx \botherref \undefined \def \botherref #1{#1}\fi
\ifx \url \undefined \def \url#1{\textsf{#1}}\fi
\ifx \bchapter \undefined \def \bchapter#1{#1}\fi
\ifx \bbook \undefined \def \bbook#1{#1}\fi
\ifx \bcomment \undefined \def \bcomment#1{#1}\fi
\ifx \oauthor \undefined \def \oauthor#1{#1}\fi
\ifx \citeauthoryear \undefined \def \citeauthoryear#1{#1}\fi
\ifx \endbibitem  \undefined \def \endbibitem {}\fi
\ifx \bconflocation  \undefined \def \bconflocation#1{#1}\fi
\ifx \arxivurl  \undefined \def \arxivurl#1{\textsf{#1}}\fi
\csname PreBibitemsHook\endcsname

\bibitem{axelrod1981evolution}
\begin{barticle}
\bauthor{\bsnm{Axelrod}, \binits{R.}},
\bauthor{\bsnm{Hamilton}, \binits{W.D.}}:
\batitle{The evolution of cooperation}.
\bjtitle{Science}
\bvolume{211}(\bissue{4489}),
\bfpage{1390}
(\byear{1981})
\end{barticle}
\endbibitem

\bibitem{schelling1980strategy}
\begin{bbook}
\bauthor{\bsnm{Schelling}, \binits{T.C.}}:
\bbtitle{The Strategy of Conflict}.
\bpublisher{Harvard University Press},
\blocation{Cambridge}
(\byear{1980})
\end{bbook}
\endbibitem

\bibitem{ratnieks2006conflict}
\begin{barticle}
\bauthor{\bsnm{Ratnieks}, \binits{F.L.W.}},
\bauthor{\bsnm{Foster}, \binits{K.R.}},
\bauthor{\bsnm{Wenseleers}, \binits{T.}}:
\batitle{Conflict resolution in insect societies}.
\bjtitle{Annu. Rev. Ento.}
\bvolume{51}(\bissue{1}),
\bfpage{581}--\blpage{608}
(\byear{2006})
\end{barticle}
\endbibitem

\bibitem{waal2000primates}
\begin{barticle}
\bauthor{\bparticle{de} \bsnm{Waal}, \binits{F.B.M.}}:
\batitle{Primates--{A} natural heritage of conflict resolution}.
\bjtitle{Science}
\bvolume{289}(\bissue{5479}),
\bfpage{586}--\blpage{590}
(\byear{2000})
\end{barticle}
\endbibitem

\bibitem{flack2006policing}
\begin{barticle}
\bauthor{\bsnm{Flack}, \binits{J.C.}},
\bauthor{\bsnm{Girvan}, \binits{M.}},
\bauthor{\bsnm{De~Waal}, \binits{F.B.M.}},
\bauthor{\bsnm{Krakauer}, \binits{D.C.}}:
\batitle{Policing stabilizes construction of social niches in primates}.
\bjtitle{Nature}
\bvolume{439}(\bissue{7075}),
\bfpage{426}--\blpage{429}
(\byear{2006})
\end{barticle}
\endbibitem

\bibitem{melis2010human}
\begin{barticle}
\bauthor{\bsnm{Melis}, \binits{A.P.}},
\bauthor{\bsnm{Semmann}, \binits{D.}}:
\batitle{How is human cooperation different?}
\bjtitle{Phil. Trans. R. Soc. B}
\bvolume{365}(\bissue{1553}),
\bfpage{2663}--\blpage{2674}
(\byear{2010})
\end{barticle}
\endbibitem

\bibitem{rand2011dynamic}
\begin{barticle}
\bauthor{\bsnm{Rand}, \binits{D.G.}},
\bauthor{\bsnm{Arbesman}, \binits{S.}},
\bauthor{\bsnm{Christakis}, \binits{N.A.}}:
\batitle{Dynamic social networks promote cooperation in experiments with
  humans}.
\bjtitle{Proc. Natl. Acad. Sci. U.S.A.}
\bvolume{108}(\bissue{48}),
\bfpage{19193}--\blpage{19198}
(\byear{2011})
\end{barticle}
\endbibitem

\bibitem{quirk1989cooperative}
\begin{barticle}
\bauthor{\bsnm{Quirk}, \binits{P.J.}}:
\batitle{The cooperative resolution of policy conflict}.
\bjtitle{Am. Polit. Sci. Rev.}
\bvolume{83}(\bissue{3}),
\bfpage{905}--\blpage{921}
(\byear{1989})
\end{barticle}
\endbibitem

\bibitem{buchan2009globalization}
\begin{barticle}
\bauthor{\bsnm{Buchan}, \binits{N.R.}},
\bauthor{\bsnm{Grimalda}, \binits{G.}},
\bauthor{\bsnm{Wilson}, \binits{R.}},
\bauthor{\bsnm{Brewer}, \binits{M.}},
\bauthor{\bsnm{Fatas}, \binits{E.}},
\bauthor{\bsnm{Foddy}, \binits{M.}}:
\batitle{Globalization and human cooperation}.
\bjtitle{Proc. Natl. Acad. Sci. U.S.A.}
\bvolume{106}(\bissue{11}),
\bfpage{4138}
(\byear{2009})
\end{barticle}
\endbibitem

\bibitem{lerner2002some}
\begin{barticle}
\bauthor{\bsnm{Lerner}, \binits{J.}},
\bauthor{\bsnm{Tirole}, \binits{J.}}:
\batitle{Some simple economics of open source}.
\bjtitle{J. Ind. Econ.}
\bvolume{50}(\bissue{2}),
\bfpage{197}--\blpage{234}
(\byear{2002})
\end{barticle}
\endbibitem

\bibitem{rogers2011teaching}
\begin{barticle}
\bauthor{\bsnm{Rogers}, \binits{D.}},
\bauthor{\bsnm{Lingard}, \binits{L.}},
\bauthor{\bsnm{Boehler}, \binits{M.L.}},
\bauthor{\bsnm{Espin}, \binits{S.}},
\bauthor{\bsnm{Klingensmith}, \binits{M.}},
\bauthor{\bsnm{Mellinger}, \binits{J.D.}},
\bauthor{\bsnm{Schindler}, \binits{N.}}:
\batitle{Teaching operating room conflict management to surgeons: clarifying
  the optimal approach}.
\bjtitle{Med. Educ.}
\bvolume{45}(\bissue{9}),
\bfpage{939}--\blpage{945}
(\byear{2011})
\end{barticle}
\endbibitem

\bibitem{minson2011two}
\begin{barticle}
\bauthor{\bsnm{Minson}, \binits{J.A.}},
\bauthor{\bsnm{Liberman}, \binits{V.}},
\bauthor{\bsnm{Ross}, \binits{L.}}:
\batitle{Two to tango: effects of collaboration and disagreement on dyadic
  judgment}.
\bjtitle{Pers. Soc. Psychol. Bull.}
\bvolume{37}(\bissue{10}),
\bfpage{1325}--\blpage{1338}
(\byear{2011})
\end{barticle}
\endbibitem

\bibitem{castellano2009statistical}
\begin{barticle}
\bauthor{\bsnm{Castellano}, \binits{C.}},
\bauthor{\bsnm{Fortunato}, \binits{S.}},
\bauthor{\bsnm{Loreto}, \binits{V.}}:
\batitle{Statistical physics of social dynamics}.
\bjtitle{Rev. Mod. Phys.}
\bvolume{81}(\bissue{2}),
\bfpage{591}--\blpage{646}
(\byear{2009})
\end{barticle}
\endbibitem

\bibitem{helbing2010quantitative}
\begin{bbook}
\bauthor{\bsnm{Helbing}, \binits{D.}}:
\bbtitle{Quantitative Sociodynamics: Stochastic Methods and Models of Social
  Interaction Processes},
\bedition{2}nd edn.
\bpublisher{Springer},
\blocation{Berlin}
(\byear{2010})
\end{bbook}
\endbibitem

\bibitem{onnela2007structure}
\begin{barticle}
\bauthor{\bsnm{Onnela}, \binits{J.-P.}},
\bauthor{\bsnm{Saram{\"a}ki}, \binits{J.}},
\bauthor{\bsnm{Hyv{\"o}nen}, \binits{J.}},
\bauthor{\bsnm{Szab{\'o}}, \binits{G.}},
\bauthor{\bsnm{Lazer}, \binits{D.}},
\bauthor{\bsnm{Kaski}, \binits{K.}},
\bauthor{\bsnm{Kert{\'e}sz}, \binits{J.}},
\bauthor{\bsnm{Barab{\'a}si}, \binits{A.-L.}}:
\batitle{Structure and tie strengths in mobile communication networks}.
\bjtitle{Proc. Natl. Acad. Sci. U.S.A.}
\bvolume{104}(\bissue{18}),
\bfpage{7332}--\blpage{7336}
(\byear{2007})
\end{barticle}
\endbibitem

\bibitem{ratkiewicz2010characterizing}
\begin{barticle}
\bauthor{\bsnm{Ratkiewicz}, \binits{J.}},
\bauthor{\bsnm{Fortunato}, \binits{S.}},
\bauthor{\bsnm{Flammini}, \binits{A.}},
\bauthor{\bsnm{Menczer}, \binits{F.}},
\bauthor{\bsnm{Vespignani}, \binits{A.}}:
\batitle{Characterizing and modeling the dynamics of online popularity}.
\bjtitle{Phys. Rev. Lett.}
\bvolume{105}(\bissue{15}),
\bfpage{158701}
(\byear{2010})
\end{barticle}
\endbibitem

\bibitem{yasseri2013value}
\begin{barticle}
\bauthor{\bsnm{Yasseri}, \binits{T.}},
\bauthor{\bsnm{Kert{\'e}sz}, \binits{J.}}:
\batitle{Value production in a collaborative environment}.
\bjtitle{J. Stat. Phys.}
\bvolume{151}(\bissue{3--4}),
\bfpage{414}--\blpage{439}
(\byear{2013})
\end{barticle}
\endbibitem

\bibitem{mestyan2013}
\begin{barticle}
\bauthor{\bsnm{Mestyán}, \binits{M.}},
\bauthor{\bsnm{Yasseri}, \binits{T.}},
\bauthor{\bsnm{Kertész}, \binits{J.}}:
\batitle{Early prediction of movie box office success based on {Wikipedia}
  activity big data}.
\bjtitle{PLoS ONE}
\bvolume{8}(\bissue{8}),
\bfpage{71226}
(\byear{2013})
\end{barticle}
\endbibitem

\bibitem{ciampaglia2011bounded}
\begin{bchapter}
\bauthor{\bsnm{Ciampaglia}, \binits{G.}}:
\bctitle{A bounded confidence approach to understanding user participation in
  peer production systems}.
In: \beditor{\bsnm{Datta}, \binits{A.}}, \betal (eds.)
\bbtitle{Social Informatics}.
\bsertitle{Lecture Notes in Computer Science},
vol. \bseriesno{6984},
pp. \bfpage{269}--\blpage{282}.
\bpublisher{Springer},
\blocation{Berlin}
(\byear{2011})
\end{bchapter}
\endbibitem

\bibitem{yasseri2012dynamics}
\begin{barticle}
\bauthor{\bsnm{Yasseri}, \binits{T.}},
\bauthor{\bsnm{Sumi}, \binits{R.}},
\bauthor{\bsnm{Rung}, \binits{A.}},
\bauthor{\bsnm{Kornai}, \binits{A.}},
\bauthor{\bsnm{Kert{\'e}sz}, \binits{J.}}:
\batitle{Dynamics of conflicts in {Wikipedia}}.
\bjtitle{PloS ONE}
\bvolume{7}(\bissue{6}),
\bfpage{38869}
(\byear{2012})
\end{barticle}
\endbibitem

\bibitem{yasseri2012circadian}
\begin{barticle}
\bauthor{\bsnm{Yasseri}, \binits{T.}},
\bauthor{\bsnm{Sumi}, \binits{R.}},
\bauthor{\bsnm{Kert{\'e}sz}, \binits{J.}}:
\batitle{Circadian patterns of wikipedia editorial activity: A demographic
  analysis}.
\bjtitle{PLoS ONE}
\bvolume{7}(\bissue{1}),
\bfpage{30091}
(\byear{2012})
\end{barticle}
\endbibitem

\bibitem{kollock1998social}
\begin{barticle}
\bauthor{\bsnm{Kollock}, \binits{P.}}:
\batitle{Social dilemmas: {The} anatomy of cooperation}.
\bjtitle{Annu. Rev. Sociol.}
\bvolume{24}(\bissue{1}),
\bfpage{183}--\blpage{214}
(\byear{1998})
\end{barticle}
\endbibitem

\bibitem{jensen2000effect}
\begin{bchapter}
\bauthor{\bsnm{Jensen}, \binits{C.}},
\bauthor{\bsnm{Farnham}, \binits{S.D.}},
\bauthor{\bsnm{Drucker}, \binits{S.M.}},
\bauthor{\bsnm{Kollock}, \binits{P.}}:
\bctitle{The effect of communication modality on cooperation in online
  environments}.
In: \bbtitle{Proceedings of the SIGCHI Conference on Human Factors in Computing
  Systems}.
\bsertitle{CHI '00},
pp. \bfpage{470}--\blpage{477}.
\bpublisher{ACM},
\blocation{New York, NY, USA}
(\byear{2000})
\end{bchapter}
\endbibitem

\bibitem{wiki:talk}
\begin{botherref}
\oauthor{\bsnm{{Wikipedia}}}:
{Wikipedia}:Talk page guidelines.
Retrieved Feb 23, 2014, from
  http://en.wikipedia.org/wiki/Wikipedia:Talk\_page\_guidelines
(2014)
\end{botherref}
\endbibitem

\bibitem{wiki:using_talk}
\begin{botherref}
\oauthor{\bsnm{{Wikipedia}}}:
{Wikipedia}:Using talk pages.
Retrieved Feb 23, 2014, from
  http://en.wikipedia.org/wiki/Wikipedia:Using\_talk\_ pages
(2014)
\end{botherref}
\endbibitem

\bibitem{kaltenbrunner2012there}
\begin{bchapter}
\bauthor{\bsnm{Kaltenbrunner}, \binits{A.}},
\bauthor{\bsnm{Laniado}, \binits{D.}}:
\bctitle{There is no deadline: {Time} evolution of {Wikipedia} discussions}.
In: \bbtitle{Proceedings of the Eighth Annual International Symposium on Wikis
  and Open Collaboration}.
\bsertitle{WikiSym '12}.
\bpublisher{ACM},
\blocation{New York, NY, USA}
(\byear{2012})
\end{bchapter}
\endbibitem

\bibitem{deffuant2000mixing}
\begin{barticle}
\bauthor{\bsnm{Deffuant}, \binits{G.}},
\bauthor{\bsnm{Neau}, \binits{D.}},
\bauthor{\bsnm{Amblard}, \binits{F.}},
\bauthor{\bsnm{Weisbuch}, \binits{G.}}:
\batitle{Mixing beliefs among interacting agents}.
\bjtitle{Adv. Complex Syst.}
\bvolume{3}(\bissue{4}),
\bfpage{87}--\blpage{98}
(\byear{2000})
\end{barticle}
\endbibitem

\bibitem{torok2013opinions}
\begin{barticle}
\bauthor{\bsnm{T{\"o}r{\"o}k}, \binits{J.}},
\bauthor{\bsnm{I{\~n}iguez}, \binits{G.}},
\bauthor{\bsnm{Yasseri}, \binits{T.}},
\bauthor{\bsnm{San~Miguel}, \binits{M.}},
\bauthor{\bsnm{Kaski}, \binits{K.}},
\bauthor{\bsnm{Kert{\'e}sz}, \binits{J.}}:
\batitle{Opinions, conflicts, and consensus: Modeling social dynamics in a
  collaborative environment}.
\bjtitle{Phys. Rev. Lett.}
\bvolume{110}(\bissue{8}),
\bfpage{088701}
(\byear{2013})
\end{barticle}
\endbibitem

\bibitem{liancourt:dispute}
\begin{botherref}
\oauthor{\bsnm{{Wikipedia}}}:
Liancourt Rocks dispute.
Retrieved May 21, 2014, from
  http://en.wikipedia.org/wiki/Liancourt\_Rocks\_dispute
(2014)
\end{botherref}
\endbibitem

\bibitem{axelrod1997dissemination}
\begin{barticle}
\bauthor{\bsnm{Axelrod}, \binits{R.}}:
\batitle{The dissemination of culture. a model with local convergence and
  global polarization}.
\bjtitle{J. Conflict Resolut.}
\bvolume{41}(\bissue{2}),
\bfpage{203}--\blpage{226}
(\byear{1997})
\end{barticle}
\endbibitem

\bibitem{lorenz2007continuous}
\begin{barticle}
\bauthor{\bsnm{Lorenz}, \binits{J.}}:
\batitle{Continuous opinion dynamics under bounded confidence: {A} survey}.
\bjtitle{Int. J. Mod. Phys. C}
\bvolume{18}(\bissue{12}),
\bfpage{1819}--\blpage{1838}
(\byear{2007})
\end{barticle}
\endbibitem

\bibitem{sznajd2005left}
\begin{barticle}
\bauthor{\bsnm{Sznajd-Weron}, \binits{K.}},
\bauthor{\bsnm{Sznajd}, \binits{J.}}:
\batitle{Who is left, who is right?}
\bjtitle{Physica A}
\bvolume{351}(\bissue{2}),
\bfpage{593}--\blpage{604}
(\byear{2005})
\end{barticle}
\endbibitem

\bibitem{wojcieszak2009underlies}
\begin{barticle}
\bauthor{\bsnm{Wojcieszak}, \binits{M.}},
\bauthor{\bsnm{Price}, \binits{V.}}:
\batitle{What underlies the false consensus effect? {How} personal opinion and
  disagreement affect perception of public opinion}.
\bjtitle{Int. J. Public Opin. R.}
\bvolume{21}(\bissue{1}),
\bfpage{25}--\blpage{46}
(\byear{2009})
\end{barticle}
\endbibitem

\bibitem{morrison2011socially}
\begin{barticle}
\bauthor{\bsnm{Morrison}, \binits{K.R.}},
\bauthor{\bsnm{Matthes}, \binits{J.}}:
\batitle{Socially motivated projection: {Need} to belong increases perceived
  opinion consensus on important issues}.
\bjtitle{Eur. J. Soc. Psychol.}
\bvolume{41}(\bissue{6}),
\bfpage{707}--\blpage{719}
(\byear{2011})
\end{barticle}
\endbibitem

\bibitem{weisbuch2002meet}
\begin{barticle}
\bauthor{\bsnm{Weisbuch}, \binits{G.}},
\bauthor{\bsnm{Deffuant}, \binits{G.}},
\bauthor{\bsnm{Amblard}, \binits{F.}},
\bauthor{\bsnm{Nadal}, \binits{J.-P.}}:
\batitle{Meet, discuss, and segregate!}
\bjtitle{Complexity}
\bvolume{7}(\bissue{3}),
\bfpage{55}--\blpage{63}
(\byear{2002})
\end{barticle}
\endbibitem

\bibitem{ben2000multiscaling}
\begin{barticle}
\bauthor{\bsnm{Ben-Naim}, \binits{E.}},
\bauthor{\bsnm{Krapivsky}, \binits{P.L.}}:
\batitle{Multiscaling in inelastic collisions}.
\bjtitle{Phys. Rev. E}
\bvolume{61}(\bissue{1}),
\bfpage{5}--\blpage{8}
(\byear{2000})
\end{barticle}
\endbibitem

\bibitem{baldassarri2002influence}
\begin{barticle}
\bauthor{\bsnm{Baldassarri}, \binits{A.}},
\bauthor{\bsnm{Marini Bettolo~Marconi}, \binits{U.}},
\bauthor{\bsnm{Puglisi}, \binits{A.}}:
\batitle{Influence of correlations on the velocity statistics of scalar
  granular gases}.
\bjtitle{Europhys. Lett.}
\bvolume{58},
\bfpage{14}
(\byear{2002})
\end{barticle}
\endbibitem

\bibitem{ben2003bifurcations}
\begin{barticle}
\bauthor{\bsnm{Ben-Naim}, \binits{E.}},
\bauthor{\bsnm{Krapivsky}, \binits{P.L.}},
\bauthor{\bsnm{Redner}, \binits{S.}}:
\batitle{Bifurcations and patterns in compromise processes}.
\bjtitle{Physica D}
\bvolume{183}(\bissue{3-4}),
\bfpage{190}--\blpage{204}
(\byear{2003})
\end{barticle}
\endbibitem

\bibitem{laguna2004minorities}
\begin{barticle}
\bauthor{\bsnm{Laguna}, \binits{M.F.}},
\bauthor{\bsnm{Abramson}, \binits{G.}},
\bauthor{\bsnm{Zanette}, \binits{D.H.}}:
\batitle{Minorities in a model for opinion formation}.
\bjtitle{Complexity}
\bvolume{9}(\bissue{4}),
\bfpage{31}--\blpage{36}
(\byear{2004})
\end{barticle}
\endbibitem

\bibitem{porfiri2007decline}
\begin{barticle}
\bauthor{\bsnm{Porfiri}, \binits{M.}},
\bauthor{\bsnm{Bollt}, \binits{E.M.}},
\bauthor{\bsnm{Stilwell}, \binits{D.J.}}:
\batitle{Decline of minorities in stubborn societies}.
\bjtitle{Eur. Phys. J. B}
\bvolume{57}(\bissue{4}),
\bfpage{481}--\blpage{486}
(\byear{2007})
\end{barticle}
\endbibitem

\bibitem{hegselmann2002opinion}
\begin{barticle}
\bauthor{\bsnm{Hegselmann}, \binits{R.}},
\bauthor{\bsnm{Krause}, \binits{U.}}:
\batitle{Opinion dynamics and bounded confidence models, analysis, and
  simulation}.
\bjtitle{J. Artif. Soc. Soc. Simul.}
\bvolume{5}(\bissue{3}),
\bfpage{2}
(\byear{2002})
\end{barticle}
\endbibitem

\bibitem{fortunato2005vector}
\begin{barticle}
\bauthor{\bsnm{Fortunato}, \binits{S.}},
\bauthor{\bsnm{Latora}, \binits{V.}},
\bauthor{\bsnm{Pluchino}, \binits{A.}},
\bauthor{\bsnm{Rapisarda}, \binits{A.}}:
\batitle{Vector opinion dynamics in a bounded confidence consensus model}.
\bjtitle{Int. J. Mod. Phys. C}
\bvolume{16}(\bissue{10}),
\bfpage{1535}--\blpage{1551}
(\byear{2005})
\end{barticle}
\endbibitem

\bibitem{jacobmeier2005multidimensional}
\begin{barticle}
\bauthor{\bsnm{Jacobmeier}, \binits{D.}}:
\batitle{Multidimensional consensus model on a {B}arab{\'a}si--{A}lbert
  network}.
\bjtitle{Int. J. Mod. Phys. C}
\bvolume{16}(\bissue{4}),
\bfpage{633}--\blpage{646}
(\byear{2005})
\end{barticle}
\endbibitem

\bibitem{lorenz2008fostering}
\begin{bchapter}
\bauthor{\bsnm{Lorenz}, \binits{J.}}:
\bctitle{Fostering consensus in multidimensional continuous opinion dynamics
  under bounded confidence}.
In: \bbtitle{{Managing Complexity: Insights, Concepts, Applications}},
pp. \bfpage{321}--\blpage{334}.
\bpublisher{Springer},
\blocation{Berlin}
(\byear{2008})
\end{bchapter}
\endbibitem

\bibitem{gonzalez2010spontaneous}
\begin{barticle}
\bauthor{\bsnm{Gonz{\'a}lez-Avella}, \binits{J.C.}},
\bauthor{\bsnm{Cosenza}, \binits{M.G.}},
\bauthor{\bsnm{Egu{\'\i}luz}, \binits{V.M.}},
\bauthor{\bsnm{San~Miguel}, \binits{M.}}:
\batitle{Spontaneous ordering against an external field in non-equilibrium
  systems}.
\bjtitle{New J. Phys.}
\bvolume{12},
\bfpage{013010}
(\byear{2010})
\end{barticle}
\endbibitem

\bibitem{lorenz2010heterogeneous}
\begin{barticle}
\bauthor{\bsnm{Lorenz}, \binits{J.}}:
\batitle{Heterogeneous bounds of confidence: {Meet}, discuss and find
  consensus!}
\bjtitle{Complexity}
\bvolume{15}(\bissue{4}),
\bfpage{43}--\blpage{52}
(\byear{2010})
\end{barticle}
\endbibitem

\bibitem{kimmons2011understanding}
\begin{barticle}
\bauthor{\bsnm{Kimmons}, \binits{R.}}:
\batitle{Understanding collaboration in {Wikipedia}}.
\bjtitle{First Monday}
\bvolume{16},
\bfpage{12}
(\byear{2011})
\end{barticle}
\endbibitem

\bibitem{wiki:war}
\begin{botherref}
\oauthor{\bsnm{{Wikipedia}}}:
{Wikipedia}:Edit warring.
Retrieved Feb 23, 2014, from
  http://en.wikipedia.org/wiki/Wikipedia:Edit\_warring
(2014)
\end{botherref}
\endbibitem

\bibitem{sumi2011edit}
\begin{bchapter}
\bauthor{\bsnm{Sumi}, \binits{R.}},
\bauthor{\bsnm{Yasseri}, \binits{T.}},
\bauthor{\bsnm{Rung}, \binits{A.}},
\bauthor{\bsnm{Kornai}, \binits{A.}},
\bauthor{\bsnm{Kert\'esz}, \binits{J.}}:
\bctitle{Edit wars in {Wikipedia}}.
In: \bbtitle{Privacy, Security, Risk and Trust (PASSAT), 2011 IEEE Third
  International Conference on and 2011 IEEE Third International Conference on
  Social Computing (SocialCom)},
pp. \bfpage{724}--\blpage{727}
(\byear{2011})
\end{bchapter}
\endbibitem

\bibitem{sepehri2012leveraging}
\begin{bchapter}
\bauthor{\bsnm{Sepehri~Rad}, \binits{H.}},
\bauthor{\bsnm{Makazhanov}, \binits{A.}},
\bauthor{\bsnm{Rafiei}, \binits{D.}},
\bauthor{\bsnm{Barbosa}, \binits{D.}}:
\bctitle{Leveraging editor collaboration patterns in {Wikipedia}}.
In: \bbtitle{Proceedings of the 23rd ACM Conference on Hypertext and Social
  Media}.
\bsertitle{HT '12},
pp. \bfpage{13}--\blpage{22}.
\bpublisher{ACM},
\blocation{New York, NY, USA}
(\byear{2012})
\end{bchapter}
\endbibitem

\bibitem{rivest1992md5}
\begin{botherref}
\oauthor{\bsnm{Rivest}, \binits{R.L.}}:
The md5 message-digest algorithm.
Internet Request for Comments,
1321
(1992)
\end{botherref}
\endbibitem

\bibitem{halfaker2011bite}
\begin{bchapter}
\bauthor{\bsnm{Halfaker}, \binits{A.}},
\bauthor{\bsnm{Kittur}, \binits{A.}},
\bauthor{\bsnm{Riedl}, \binits{J.}}:
\bctitle{Don't bite the newbies: How reverts affect the quantity and quality of
  wikipedia work}.
In: \bbtitle{Proceedings of the 7th International Symposium on Wikis and Open
  Collaboration}.
\bsertitle{WikiSym '11},
pp. \bfpage{163}--\blpage{172}.
\bpublisher{ACM},
\blocation{New York, NY, USA}
(\byear{2011})
\end{bchapter}
\endbibitem

\bibitem{yasseri2014controversial}
\begin{bchapter}
\bauthor{\bsnm{Yasseri}, \binits{T.}},
\bauthor{\bsnm{Spoerri}, \binits{A.}},
\bauthor{\bsnm{Graham}, \binits{M.}},
\bauthor{\bsnm{Kertész}, \binits{J.}}:
\bctitle{The most controversial topics in {Wikipedia}: A multilingual and
  geographical analysis}.
In: \beditor{\bsnm{Fichman}, \binits{P.}},
\beditor{\bsnm{Hara}, \binits{N.}} (eds.)
\bbtitle{Global {Wikipedia}: International and Cross-cultural Issues in Online
  Collaboration}.
\bpublisher{Scarecrow Press},
\blocation{Lanham, Md}
(\byear{2014})
\end{bchapter}
\endbibitem

\bibitem{wiki:ban}
\begin{botherref}
\oauthor{\bsnm{{Wikipedia}}}:
{Wikipedia}:Banning policy.
Retrieved Feb 23, 2014, from
  http://en.wikipedia.org/wiki/Wikipedia:Banning\_policy
(2014)
\end{botherref}
\endbibitem

\end{thebibliography}


\end{backmatter}
\end{document}